\documentclass[a4paper,11pt]{article}
\pdfoutput=1 

\usepackage{jcappub} 

\usepackage[T1]{fontenc} 

\usepackage{graphicx}
\usepackage{caption}
\usepackage{subcaption}
\usepackage{dcolumn}
\usepackage{bm}
\usepackage{hyperref}
\usepackage{color}
\usepackage{amsmath}
\usepackage{acronym}
\usepackage{xspace}
\usepackage{soul}

\newcommand{\gwcosmo}{\textsc{gwcosmo}\xspace}
\newcommand{\icarogw}{\textsc{Icarogw}\xspace}

\acrodef{H0}[$H_0$]{the Hubble constant}
\acrodef{LVC}[LVC]{LIGO and Virgo Collaborations}
\acrodef{GW}[GW]{gravitational wave}
\acrodef{EM}[EM]{electromagnetic}
\acrodef{CMB}[CMB]{cosmic microwave background}
\acrodef{SN}[SN]{supernovae}
\acrodef{MDC}[MDA]{mock data analysis}
\acrodefplural{MDC}[MDAs]{mock data analyses}
\acrodef{BNS}[BNS]{binary neutron star}
\acrodef{BBH}[BBH]{binary black hole}
\acrodef{LCDM}[$\Lambda$CDM]{$\Lambda$-cold-dark-matter}
\acrodef{SNR}[SNR]{signal-to-noise ratio}
\acrodef{F2Y}[F2Y]{First Two Years}
\acrodef{GRB}[GRB]{gamma-ray burst}
\acrodef{SH0ES}[SH0ES]{Supernovae, $H_0$, for the Equation of State of Dark energy}
\acrodef{O1}[O1]{first observing run}
\acrodef{O2}[O2]{second observing run}
\acrodef{O3}[O3]{third observing run}
\acrodef{NSBH}[NSBH]{neutron star - black hole pair}
\acrodef{CBC}[CBC]{compact binary coalescence}
\acrodef{O2H0}[O2-$H_0$]{O2 Hubble constant}
\acrodef{PSD}[PSD]{power-spectral-density}
\acrodefplural{PSD}[PSDs]{power-spectral-densities}
\acrodef{RA}[RA]{right ascension}
\acrodef{dec}[dec]{declination}
\acrodef{dl}[$d_L$]{luminosity distance}
\acrodef{LOS}[LOS]{line-of-sight}
\acrodef{KDE}[KDE]{kernel density estimate}
\acrodef{FAR}[FAR]{false alarm rate}
\acrodef{IFAR}[IFAR]{inverse false alarm rate}
\acrodef{PE}[PE]{parameter estimation}
\acrodef{DES}[DES]{Dark Energy Survey}
\acrodef{SDSS}[SDSS]{Sloan Digital Sky Survey}
\acrodef{GLADE}[GLADE]{Galaxy List for the Advanced Detector Era}
\acrodef{GWENS}[GWENS]{Gravitational Wave Events in Sloan}
\acrodef{LSC}[LSC]{LIGO Scientific Collaboration}
\acrodef{mth}[$m_\text{th}$]{apparent magnitude threshold}
\acrodef{GWTC-3}[GWTC-3]{the third Gravitational-Wave Transient Catalogue}
\acrodef{SFR}[SFR]{star formation rate}

\title{\boldmath Testing the nature of gravitational wave propagation using dark sirens and galaxy catalogues}

\author[a]{Anson Chen, }
\author[a,b]{Rachel Gray, }
\author[a,c]{Tessa Baker}

\affiliation[a]{Department of Physics and Astronomy, Queen Mary University of London, Mile End Road, London, E1 4NS, United Kingdom}
\affiliation[b]{SUPA, University of Glasgow, Glasgow, G12 8QQ, United Kingdom}
\affiliation[c]{Institute of Cosmology and Gravitation, University of Portsmouth, Burnaby Road, Portsmouth PO1 3FX, UK}

 \emailAdd{anson.chen@ligo.org}
 \emailAdd{rachel.gray@ligo.org}
 \emailAdd{tessa.baker@ligo.org}

\date{\today}

\abstract{The dark sirens method enables us to use gravitational wave events \textit{without} electromagnetic counterparts as tools for cosmology and tests of gravity. Furthermore, the dark sirens analysis code \gwcosmo can now robustly account for information coming from both galaxy catalogues and the compact object mass distribution. We present here an extension of the \gwcosmo code and methodology to constrain parameterized deviations from General Relativity that affect the propagation of gravitational waves. We show results of our analysis using data from the GWTC-3 gravitational wave catalogues, in preparation for application to the O4 observing run. After testing our pipelines using the First Two Years mock data set, we reanalyse 46 events from GWTC-3, and combine the posterior for BBH and NSBH sampling results for the first time. We obtain joint constraints on $H_0$ and parameterized deviations from General Relativity in the Power Law + Peak BBH population model. With increased galaxy catalogue support in the future, our work sets the stage for dark sirens to become a powerful tool for testing gravity.}

\begin{document}
\maketitle
\flushbottom


\section{Introduction}

\noindent  
{As the imprint of a propagating gravitational field perturbation, gravitational waves (GWs) are the most natural tool with which to test the nature of gravity. The nature of their generation and subsequent propagation means that they uniquely access both the strong-field regime of gravity near compact objects, as well as the weak-field regime of gravity on cosmological scales \cite{PhysRevD.103.122002, LIGOScientific:2021sio}. Furthermore, unlike many electromagnetic (EM) observables, they carry direct information about the luminosity distance of their source. When combined with redshift information, we can therefore use gravitational wave data to probe the luminosity distance-redshift relation -- and hence key cosmological parameters like the Hubble constant -- in a manner distinct to traditional methods such as the Cosmic Microwave Background or the local distance ladder \cite{Verde2019,DIVALENTINO2021102605}. Some departures from General Relativity (GR) will also impact the luminosity distance-redshift relation, and hence can be tested in this way.

For both tests of gravity and measurements of $H_0$, the presence of an EM counterpart to a GW source -- a ``bright siren'' event first introduced by \cite{Schutz1986} -- provides the most direct and (usually) powerful constraints. This was seen most spectacularly with binary neutron star (BNS) merger GW170817 and its EM counterpart \cite{PhysRevLett.119.161101,Abbott_2017_cT}. Here, the localization of the accompanying kilonova allowed the galaxy NGC4993 to be uniquely identified as the host galaxy of GW170817 \cite{LIGOScientific:2017ync}. With an additional correction for the peculiar velocity of NGC4993, from this single event alone the Hubble constant was estimated to be $H_0=70.0^{+12.0}_{-8.0}~{\rm km}~{\rm s}^{-1}{\rm Mpc}^{-1}$ \cite{Abbott2017_H0}. The implications of GW170817 for tests of gravity were even more impactful. The arrival of the GRB counterpart just $\sim 1.7$s after the GW merger signal constrained the propagation speed of GWs ($c_T$) in the frequency band of terrestrial detectors\footnote{A caveat is that an effective field approach to modified gravity still allows a deviation of $c_T$ from $c$ at an energy scale lower than the terrestrial detection band \cite{deRham:2018red}. This scenario can be probed with massive BBHs by the Laser Interferometer Space Antenna (LISA) alone \cite{LISACosmologyWorkingGroup:2022wjo}, or with stellar origin BBHs by joint detection of LVK and LISA \cite{Baker:2022eiz,Harry:2022zey}, or possibly by pulsar timing arrays.} at the level $|1-c_T/c|\leqslant 3\times10^{-15}$ \cite{Abbott_2017_cT}. This led to stringent constraints on various modified gravity models, such as scalar-tensor and vector-tensor theories\cite{Baker:2017hug,PhysRevLett.119.251304,PhysRevLett.119.251302}, Born-Infeld gravity\cite{PhysRevD.97.084011} and parity-violating gravity\cite{PhysRevD.98.124018}. 

However, so far bright sirens have proved relatively elusive. With current broad uncertainties on the underlying event rates, it remains unclear to what extent we can rely on them as cosmological tools. This has led to the development of an alternative set of techniques for GW cosmology that do \textit{not} rely on EM counterparts, known as dark siren methods. These techniques instead employ galaxy catalogues, together with features in the mass spectrum of compact objects, to supply redshift information about the possible hosts of GW sources. Under these methods, the redshift of a GW host galaxy cannot be uniquely identified; instead the galaxy catalogue is used to construct a prior distribution for the redshift along specific lines of sight (in a pixellated fashion \cite{Gray2022}). This prior is combined with GW data in a hierarchical Bayesian formalism \cite{PhysRevD.86.043011,chen2018,Fishbach_2019,PhysRevD.101.122001} to infer cosmological parameters. The population model of compact objects also informs this result because, assuming a known distribution of compact object masses in the source frame, this is related to the distribution that GW detectors actually measure (in `detector frame') by factors of $(1+z)$. Hence for any distinct features in the mass distribution we can effectively compare their assumed and measured locations to infer information about source redshifts, again in a statistical sense only.

To date, the predominant focus of dark sirens development has been to constrain the Hubble constant (but see also \cite{Leyde_2022,Finke:2021aom}). The primary goal of this paper is to introduce an extension of a key dark sirens tool, described below, which allows the method to test extensions of GR impacting the luminosity distance-redshift relation, i.e. one of the most generic effects of cosmologically-motivated modified gravity theories. With the fourth observing run of the LIGO-Virgo-KAGRA (LVK) collaboration underway, we ready our tools for application to the increasing stack of GW detections, to yield (we anticipate) increasingly precise results. 
 
At present, two sophisticated code packages have been independently developed to measure cosmological parameters with the bright siren and/or the dark siren methods: \gwcosmo \cite{Gray2022,Gray:2023wgj} and \icarogw \cite{mastrogiovanni2023novel,Mastrogiovanni:2023zbw}. Both pipelines are now capable of drawing on information from both the statistical host galaxy distribution from galaxy catalogues, and also via constraining compact object population models and merger rate redshift evolution models. The ability to use information from both galaxy catalogues and the population model simultaneously is an important recent development, detailed in \cite{Gray:2023wgj}, which ensures a robust $H_0$ result\footnote{Previously, the first version of \icarogw \cite{2021PhRvD.104f2009M} constrained $H_0$, population and merger rate models without utilising galaxy catalogue information; meanwhile \gwcosmo v1.0.0 \cite{Gray2022} measured $H_0$ using galaxy catalogue information, but fixing the population and merger rate model.}. 

After the O3 observation run \cite{LIGOScientific:2021djp}, 46 dark sirens in GWTC-3 with signal-to-noise ratios (SNRs) larger than 11 were selected for statistical host galaxy identification with the GLADE+ galaxy catalogue \cite{Gray2022} using \gwcosmo v1.0.0 \cite{GWTC-3:cosmology}. These yield a measurement of $H_0=67^{+13}_{-12}~{\rm km}~{\rm s}^{-1}{\rm Mpc}^{-1}$. Combining the dark siren result with the bright siren GW170817 yields $H_0=68^{+8}_{-6}~{\rm km}~{\rm s}^{-1}{\rm Mpc}^{-1}$. Meanwhile the analysis with 42 BBHs using \icarogw 2.0 gives $H_0=71^{+35}_{-30}~{\rm km}~{\rm s}^{-1}{\rm Mpc}^{-1}$ with the assumption of the compact object binary merger rate proportional to the galaxy luminosity, and $H_0=43^{+48}_{-18}~{\rm km}~{\rm s}^{-1}{\rm Mpc}^{-1}$ without the assumption \cite{mastrogiovanni2023novel}. Although these measurements are not more precise than those from EM probes, they serve as an early step on a pathway towards and independent and competitive probe of $H_0$. We note here that there exists a third approach to constraining $H_0$ with GW data, by fully cross-correlating GW events and galaxy clustering \cite{Mukherjee:2022afz,Bera:2020jhx,PhysRevD.93.083511}.

We wish to capitalise on these early successes of the dark sirens methodology, and likewise apply them to test the laws of gravity. The purpose of this paper is to describe the extension of the \gwcosmo software to achieve precisely this. By reanalysing GWTC-3 events and First Two Years mock data, we demonstrate the constraints on deviations obtained by \gwcosmo and lay the groundwork for future tests of gravity at a cosmic scale in O4 that have not been possible in previous observing runs. The inclusion of similar modified gravity effects in the \icarogw code is described in \cite{Leyde_2022, mastrogiovanni2023novel}, whilst an independent code pipeline and results also appears in \cite{Finke:2021aom}.

The paper is constructed in five sections. In Section \ref{sec:method} we review possible modifications to the GW luminosity distance outside of GR, and introduce three classes of parameterization for this phenomenon. In Section \ref{sec:gwcosmo} we revisit the theoretical framework of \gwcosmo and develop its extension to features beyond GR. Section \ref{sec:mock_data} and Section \ref{sec:GWTC-3} show the reanalysis of mock bright siren data and GWTC-3 events respectively, now including modified gravity parameters with the latest version of \gwcosmo. We conclude and discuss further lines of development in Section \ref{sec:conclusions}.

\section{GW Propagation Beyond GR\label{sec:method}}

\noindent 
A common feature of gravity theories beyond GR is a modification to the effective friction term that impacts GW propagation. In theories with this feature, in the Jordan frame (where matter is minimally coupled to the metric), the modified GW propagation equation can be written as \cite{PhysRevD.97.104066, PhysRevD.98.023510, PhysRevD.97.104037, Belgacem_2019, Baker_2021, Finke:2021aom}
\begin{equation}
    \tilde{h}_A'' + 2{\cal H}[1-\delta(\eta)]\tilde{h}_A'+c^2k^2\tilde{h}_A = 0,
\label{eq:GW_propagation}
\end{equation}
where $\tilde{h}(\eta,k)$ is the GW amplitude in Fourier space, subscript $A$ denotes the $+$ or $\times$ polarization of GWs, $\eta$ is conformal time, $'$ denotes the derivative with respect to $\eta$, and ${\cal H}=a'/a$ is the conformal Hubble parameter. Here $\delta(\eta)$ is the modification to the friction term introduced by modified gravity theories, where GR is recovered when $\delta(\eta)=0$. It is shown in Sec. 4.1.4 of \cite{Maggiore:2007ulw} that the solution to Eq. \eqref{eq:GW_propagation} in GR gives $\tilde{h}(\eta,k) \propto 1/d_L^{\rm EM}(z)$, where $d_L^{\rm EM}(z)$ is the EM luminosity distance defined by 
\begin{equation}
    d_L^{\rm EM}(z) = (1+z) \int_0^z \frac{c~d\tilde{z}}{H_0\sqrt{\Omega_{m,0}(1+\tilde{z})^3+\Omega_{\Lambda,0}}}
    \label{eq:d_EM}
\end{equation}
in the late-time universe. $\Omega_{m,0}$ and $\Omega_{\Lambda,0}$ are the matter density today and the dark energy density today respectively. When $\delta(\eta) \neq 0$, it is shown in \cite{PhysRevD.97.104066} that one can absorb the changes to the GW propagation equation by defining a new effective conformal Hubble factor. To do this, we introduce a new scale factor $\tilde{a}$ such that 
\begin{equation}
    \frac{\tilde{a}'}{\tilde{a}} = {\cal H}[1-\delta(\eta)],
    \label{eq:new_scale_factor}
\end{equation}
When expressed in terms of the new scale factor, the friction term now has the same form as in GR. From the solution to the transformed equation we obtain a new quantity called the GW luminosity distance $d_L^{\rm GW}(z)$, which is related to $d_L^{\rm EM}(z)$ by 
\begin{equation}
    d_L^{\rm GW}(z) = \frac{a(z)}{\tilde{a}(z)} d_L^{\rm EM}(z),
\end{equation}
In a non-GR theory, we now have $\tilde{h}(\eta,k) \propto 1/d_L^{\rm GW}(z)$. Eq. (\ref{eq:new_scale_factor}) can be written as 
\begin{equation}
    \left(\log\frac{a}{\tilde{a}} \right)' = \delta(\eta){\cal H}(\eta).
    \label{eq: transform_scale_factor}
\end{equation}
Integrating Eq. (\ref{eq: transform_scale_factor}) allows us to re-express the relation between $d_L^{\rm GW}(z)$ and $d_L^{\rm EM}(z)$ as:
\begin{equation}
    d_L^{\rm GW}(z) = d_L^{\rm EM}(z) \exp \left\{ -\int^z_0 \frac{dz'}{1+z'}\,\delta(z') \right\}.
    \label{eq:dgw_dL}
\end{equation}
The exact form of $\delta(z)$ depends on the specific gravity model under consideration (some subtleties related to this parameterisation are discussed in \cite{PhysRevLett.130.231401}).We can also choose to parameterise the entire term in curly brackets above in some sensible manner, which we demonstrate below. In the next section we present the analysis of three models/parameterizations implemented in \gwcosmo.

\subsection{$(\Xi_0, n)$ parameterization}
\label{subsec:Xi0n}

\noindent As mentioned above, there exist many gravity theories that can introduce a non-standard friction term in the GW propagation equation. Under these circumstances, rather than work with a large number of individual models, it can be efficient to instead work with a general parameterized form of the ratio $d_L^{\rm GW}(z)/d_L^{\rm EM}(z)$. Constraints on parameters belonging to specific gravity models are then obtainable via a mapping from the parameterised constraint.

 A widely-adopted and simple parameterization for the ratio of $d_L^{\rm GW}(z)/d_L^{\rm EM}(z)$ is given by \cite{PhysRevD.98.023510}
\begin{equation}
    \frac{d_L^{\rm GW}(z)}{d_L^{\rm EM}(z)} = \Xi_0 + \frac{1-\Xi_0}{(1+z)^n},
    \label{eq:dgw_Xi}
\end{equation}
where $\Xi_0$ and $n$ are the parameters controlling the variation from GR. This parameterization was originally obtained by fitting to a class of non-local gravity models in \cite{PhysRevD.98.023510,Belgacem:2017ihm,Belgacem:2017cqo},
however, its form can be motivated on more general grounds as we describe below momentarily. As a result, this parameterization has been widely adopted in testing GR with dark sirens, for example in \cite{Finke:2021aom,Leyde_2022,mastrogiovanni2023novel,Mancarella:2021ecn,Mancarella:2022cnu}, and with other techniques such as using BNS mass distribution \cite{Finke:2021eio} and strongly lensed GW events \cite{Finke:2021znb}. Some mappings of $\Xi_0$ and $n$ to parameters in exact gravity models are listed in Table 1 of \cite{Belgacem_2019}.} 

The GR limit of Eq.(\ref{eq:dgw_Xi}) corresponds to $\Xi_0=1$, so that the GW distance is the same as the EM luminosity distance. This limit is also recovered for $z\sim 0$, largely irrespective of $\Xi_0$. Conversely, when $z\gg1$ the ratio $d_L^{\rm GW}(z)/d_L^{\rm EM}(z)$ approaches a constant set by the value of $\Xi_0$. This is motivated by the hypothesis that any deviation from GR predominantly affects the late universe where dark energy dominates, so $d_L^{\rm GW}(z)/d_L^{\rm EM}(z)$ deviates from 1 as redshift increases under effects beyond GR, and approaches constant at high redshift since the deviation from GR is suppressed in early universe.

 The value of $n$ determines how sharply the transition of the ratio from 1 to $\Xi_0$ take place as redshift increases. In the low redshift limit $z\ll1$, the ratio can be approximated at the linear order in $z$:
\begin{equation}
\frac{d_L^{\rm GW}(z)}{d_L^{\rm EM}(z)} = 1-zn(1-\Xi_0)+O(z^2),
\label{eq:Xi_lowz}
\end{equation}
which is often useful in the regime of the local universe. When $n$ is fixed as a parameter, its fiducial value is often taken to be $n=1.91$, the value predicted by the RT non-local gravity model when measuring $\Xi_0$ \cite{Finke:2021aom}. This may seem a somewhat arbitrary choice, but it will not significantly impact the results of this work (typically we will allow $n$ to vary as a free parameter with broad priors anyway). Note that GR is also recovered as $n$ approaches 0, meaning we can potentially expect to see some degeneracy between constraints on $\Xi_0$ and $n$, if the data are consistent with GR.

In Fig. \ref{fig:dgw_z} we plot the ratio of $d_L^{\rm GW}(z)/d_L^{\rm EM}(z)$ for some different values of $\Xi_0$. The ratio is greater than 1 when $\Xi_0>1$ and lower than 1 when $\Xi_0<1$, for $n \neq 0$.  

\subsection{Horndeski class parameterization}

\noindent In addition to the generic $(\Xi_0,n)$ parameterization, we also implement a specific parameterization originating from the Horndeski family of extensions to GR. The Horndeski action describes the most generic scalar-tensor theory of gravity with a second-order equation of motion \cite{Horndeski:1974wa,Deffayet:2011gz}. The observation of the luminal speed of gravitational waves by GW170817 and its EM counterpart \cite{PhysRevLett.119.161101} strongly constrains particular terms in the original Horndeski action, though see the caveats discussed in \cite{deRham:2018red}. Setting these terms to zero, the remaining reduced Horndeski action is \cite{Baker:2017hug, Ezquiaga:2017ekz, Creminelli:2017sry, Sakstein:2017xjx} 
\begin{equation}
    S = \int d^4x\sqrt{-g}\left[\frac{1}{2}M^2_{\rm eff}(\phi)R+K(\phi,X)-G_3(\phi,X)\Box\phi \right] + S_m(g_{\mu\nu},\psi_m),
\end{equation}
where $\phi$ is the scalar degree of freedom, $M_{\rm eff}$ is the effective Planck mass which evolves with $\phi$, $X$ is a kinetic term defined by $X \equiv -\bigtriangledown^\mu \phi \bigtriangledown_\mu \phi/2$, $S_m$ is the matter action, and $\psi_m$ are matter fields minimally coupled to the metric $g_{\mu\nu}$. When $M_{\rm eff}$ equals the Planck mass $M_{\rm P}$, and $K=G_3=0$, the action reduces to GR. It has been shown in the literature (e.g. \cite{Bellini_2014,Lagos_2018,Baker_2021}) that the GW propagation equation derived from the reduced Horndeski action is given by
\begin{equation}
    \tilde{h}'' + {\cal H}[2+\alpha_M(z)]\tilde{h}'+c^2k^2\tilde{h} = 0,
\end{equation}
where $\alpha_M$ is the running rate of the effective Planck mass defined by
\begin{equation}
    \alpha_M(z) \equiv \frac{d\ln(M_{\rm eff}/M_{\rm P})^2}{d\ln a}.
\end{equation}
We see from Eq. \eqref{eq:GW_propagation} that the GW friction term $\delta(z)$ is related to $\alpha_M(z)$ by $\delta(z) = -\alpha_M(z)/2$. A widely-used and simple parameterization for $\alpha_M(z)$ is that it is proportional to the fractional dark energy density \cite{Bellini_2016,PhysRevD.95.063502,PhysRevD.99.083504,PhysRevD.102.044009} as
\begin{equation}
    \alpha_M(z) = c_M \frac{\Omega_\Lambda(z)}{\Omega_{\Lambda,0}} = c_M \frac{1}{\Omega_{m,0} (1+z)^3 + \Omega_{\Lambda,0}},
\label{eq:alphaM_cM}
\end{equation}
where $c_M$ is the free parameter to be constrained, and $\Omega_\Lambda(z)$ is the fractional dark energy density. This assumption associates the deviation from GR to the growth of the dark energy density parameter. On one hand, this seems a plausible assumption if a cosmological modified gravity theory fulfils its principle motivation to replace dark energy. On the other, it has been argued that Eq. \eqref{eq:alphaM_cM} is not representative of the true evolution of mainstream gravity theories \cite{Linder:2016wqw}. We will use Eq. \eqref{eq:alphaM_cM} as a general agnostic expression; a different parameterization can be straightforwardly replaced in our calculations. For instance another way of parameterization can be found in \cite{DAgostino:2019hvh}.

With the relation $\delta(z) = -\alpha_M(z)/2$, one obtains by substituting the above parameterization into Eq. \eqref{eq:dgw_dL} that
\begin{equation}
    d_L^{\rm GW}(z) = d_L^{\rm EM}(z) \exp \left\{\frac{c_M}{2\Omega_{\Lambda,0}} \ln \frac{1+z}{[\Omega_{m,0}(1+z)^3+\Omega_{\Lambda,0}]^{1/3}} \right\}.
\label{eq:dgw_cM}
\end{equation}
This parameterization only depends on one free parameter, $c_M$. GR is recovered when $c_M=0$. We have $d_L^{\rm GW}(z)/d_L^{\rm EM}(z)>1$ when $c_M>0$, and $d_L^{\rm GW}(z)/d_L^{\rm EM}(z)<1$ when $c_M<0$, as shown in Fig. \ref{fig:dgw_z}.

\subsection{Extra dimensions}

\noindent Apart from scalar-tensor theories, additional spacetime dimensions are introduced in some extended theories of gravity, which can also leave an imprint on GW propagation. For instance, in Dvali-Gabadadze-Porrati (DGP) theory \cite{DVALI2000208,Antonio_Padilla_2004}, gravitational forces propagate in higher dimensions beyond 4-dimension spacetime, while ordinary matter is confined on the 3-dimension spatial brane. This leads to an effective `leakage' of the perceived gravitational forces over cosmological distance scales, which transforms into an extra damping term in the GW propagation equation. It is shown in \cite{Corman_2021} that the GW and EM luminosity distances in some theories with extra dimensions are related by
\begin{equation}
    d_L^{\rm GW}(z) = d_L^{\rm EM}(z) \left\{1+\left[\frac{d_L^{\rm EM}(z)}{(1+z)R_c} \right]^{n_D} \right\}^{\frac{D-4}{2n_D}},
    \label{eq:dgw_D}
\end{equation}
where $D$ is the number of spacetime dimensions, $R_c$ is the comoving scale which controls the transitions from the GR 4-dimensional regime at small scales to the higher-dimensional regime on large scales, and $n_D$ controls the sharpness of the transition around $R_c$. On small scales where $d_L^{\rm EM} \ll R_c$, GR is recovered and hence $d_L^{\rm GW}$ is indistinguishable from $d_L^{\rm EM}$. On large scales where $d_L^{\rm EM} \gg R_c$, the ratio of $d_L^{\rm GW}/d_L^{\rm EM}$ grows with the comoving distance to the power of $(D-4)/2$. Constraints on $D$ with bright and dark sirens have been studied in \cite{Corman_2021,Leyde_2022}.

\begin{figure}
    \centering
    \includegraphics[width=0.6\textwidth]{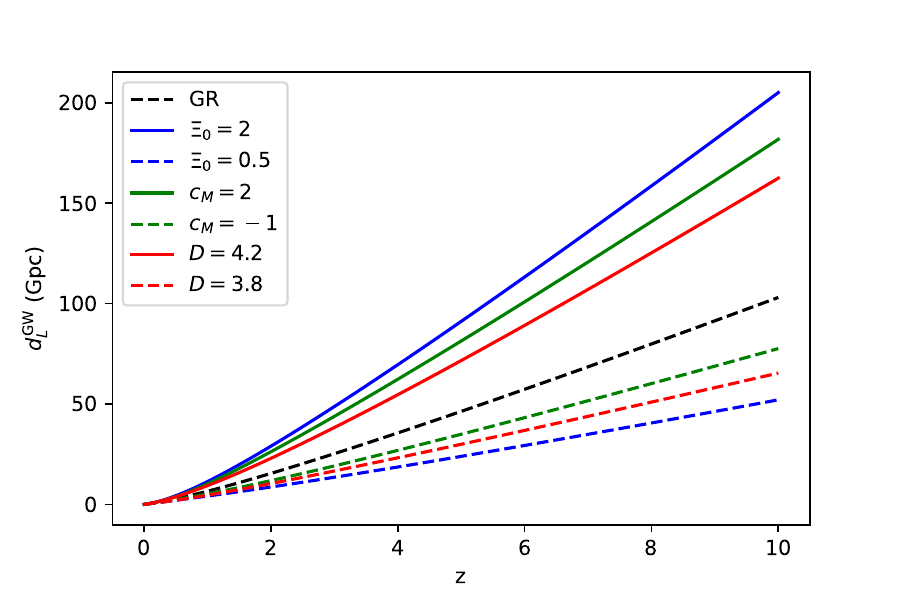}
    \caption{GW luminosity distances in the three modified gravity models analysed in this paper compared to GR. We fix $n=1.91$ in the $(\Xi_0,n)$ parameterization, and $R_c=100$ Mpc in the extra-dimension model. }
    \label{fig:dgw_z}
\end{figure}

The comparison between $d_L^{\rm GW}$ and $d_L^{\rm EM}$ for the three modified gravity parameterizations with a selection of typical values is shown in Fig.~\ref{fig:dgw_z}. We can see that in all of the three models, $d_L^{\rm GW}(z)/d_L^{\rm EM}(z)$ can either be larger or smaller than 1, depending on the parameter value. It starts from 1 at low redshift and deviates from 1 as redshift increases. Therefore the deviation from GR is more significant at high redshift, so at the simplest level, GW events further away from may be more helpful in testing these gravity models. As we will see in sections \ref{sec:mock_data} and \ref{sec:GWTC-3}, the reality of this statement will be complicated by the sky localisations and distance errors of each event.

\section{Modifications to \gwcosmo \label{sec:gwcosmo}}

\subsection{Bayesian framework of \gwcosmo}

\noindent The detailed Bayesian statistical framework of the \gwcosmo code is presented in \cite{PhysRevD.101.122001,Gray2022,Gray:2023wgj}, which we will briefly review in this section. Then we will show how the posterior probabilities of modified gravity parameters are built. They are included in a set of hyper-parameters $\Lambda$ to be measured with the bright siren and the dark siren methods. These hyper-parameters also include cosmological parameters like $H_0$, and parameters describing both the compact object population distribution and the merger rate redshift evolution, which will be discussed in section \ref{sec:prior_mass_rate} .

Given $N_\text{det}$ detected GW events with observation data $\{x_\text{GW}\}$, the posterior for a set of hyper-parameters $\Lambda$ is given by \cite{10.1093/mnras/stz896,Vitale2020}
\begin{equation}
\begin{aligned}
p(\Lambda|\{x_\text{GW}\},\{D_\text{GW}\},I) &\propto p(\Lambda|I) p(N_\text{det}|\Lambda,I)  \prod^{N_\text{det}}_i p(x_{\text{GW}i}|D_{\text{GW}i},\Lambda,I),
\label{eq:pLam}
\end{aligned}
\end{equation}
where $D_\text{GW}$ is a parameter representing whether a GW event is detected or not, which takes the value of 1 or 0. $I$ represents any other information or assumptions not explicitly contained in the parameter set $\Lambda$. The detection criteria of an event is that its SNR surpasses the SNR threshold we choose. The selected SNR threshold in our analysis will be discussed in the next two sections. The prior $p(\Lambda|I)$ is uniform in the analysis for all parameters in $\Lambda$. $p(N_\text{det}|\Lambda,I)$ is the probability of detecting $N_\text{det}$ events. It depends on the intrinsic astrophysical rate of GW events $R$. As discussed in \cite{Fishbach:2018edt, 10.1093/mnras/stz896}, by choosing a flat prior on $R$ in a log scale, we obtain the combined prior $p(\Lambda,R|I)\propto 1/R$, so that the dependence on $R$ is dropped out when marginalizing $p(N_\text{det}|\Lambda,I)$ over $R$. 
Beyond this $p(N_\text{det}|\Lambda,I)$ has no further dependence on hyper-parameters. With Bayes' theorem, the likelihood for obtaining a set of GW observations $x_{\text{GW}}$ can be expanded into
\begin{equation}
\begin{aligned}
\prod^{N_\text{det}}_i p(x_{\text{GW}i}|D_{\text{GW}i},\Lambda,I) &\propto \prod^{N_\text{det}}_i \dfrac{\int p(x_{\text{GW}i}|\theta,\Lambda,I)p(\theta|\Lambda,I) d\theta}{\int p(D_{\text{GW}i}|\theta,\Lambda,I)p(\theta|\Lambda,I) d\theta} \\
& \propto {\bigg[\prod^{N_\text{det}}_i\int p(x_{\text{GW}i}|\theta,\Lambda,I)p(\theta|\Lambda,I) d\theta\bigg]}{\bigg[\int p(D_\text{GW}|\theta,\Lambda,I)p(\theta|\Lambda,I) d\theta \bigg]^{-N_\text{det}}},
\end{aligned}
\end{equation}
where $\theta$ represents a parameter set describing a detected GW signal, which must be marginalised over in order to find posterior for $\Lambda$. The parameters in $\theta$ of interest to us in this work include the detector frame masses $m_1^{\rm det}$ and $m_2^{\rm det}$, redshift $z$, and sky location $\Omega$. 
$p(x_{\text{GW}i}|\theta,\Lambda,I)$ in the numerator is obtained from the posterior samples of GW signal parameters estimated for each detected event, assuming a cosmological model. Here only cosmological hyper-parameters are relevant, while population and merger rate hyper-parameters enter in $p(\theta|\Lambda,I)$.  $p(D_{\text{GW}i}|\theta,\Lambda,I)$ in the denominator is called the detection probability of GW events given a cosmology model. In \gwcosmo it is computed with a large number of injections of GW events. The injections are simulated detections of GW signals that are generated with the GW waveform injected with noises. The injections we used are generated with the IMRPhenomPv2 waveform over the sensitivity curves of specific detector network configurations such as O2, O3 or O4 set ups. Provided the selected SNR threshold, injections meeting the SNR threshold are used for computing the detection probability.

The posterior computed above is affected by the selection effects, which are the effects caused by the sensitivity of GW detectors, our choices of the population and cosmology model, prior ranges of $\theta$ and the SNR threshold in computing detection probability. If the prior range of parameters cannot cover the space that may be detected from GW events, the posterior will be biased. However, it is not always more beneficial for the prior range to be larger, because some parameters may be degenerate, so that one of the unconstrained parameters may biase constraints on another one (we will see an example of this in Section \ref{sec:GWTC-3}). In addition, the SNR threshold determines whether an event is counted as detected. It must be consistent with the SNR threshold of LVK detection, so it can't be too low. 
On the other hand, if the threshold is too high, we will have too few events to obtain a tighter constraint. Therefore the prior range and the SNR threshold need to be selected carefully.

 The probabilistic distributions of the redshift and sky location of the GW event are marginalised over in different ways for the bright siren and the dark siren methods. For a bright siren, a prior on the redshift and the sky location can be retrieved from the host galaxy identified from an EM counterpart with high accuracy. In \gwcosmo the redshift prior of a bright siren $p(z|\Lambda, I)$ is approximated by a Gaussian distribution with the mean and the standard deviation being the same as the measured redshift of the host galaxy and its redshift uncertainty. (This in principle could be generalised to non-Gaussian redshift measurements in future.) 

On the other hand, the prior on redshift and the sky location in the dark siren method is obtained with galaxy catalogue information. Without an EM counterpart, a GW event is localized within a patch of the sky that is too large to identify its host galaxy. However, galaxies recorded in catalogues within this patch of the sky are potential host galaxies that can provide redshift information of the GW source. The redshift prior of the GW source is then obtained by combining the redshift distribution of galaxies in the catalogues within the sky localization area of the event upon the prior of the comoving redshift volume. 

In \gwcosmo the redshift prior is computed for each pixel on the sky \cite{Gray2022}. The resolution in dividing the sky is determined by a parameter called nside introduced by \texttt{healpy} \cite{Zonca2019,Gorski_2005}. For larger nside, there are more pixels in the whole sky and correspondingly fewer galaxies in each pixel; this allows for more finely-grained features in across different patches in the sky, but takes takes longer to evaluate numerically. For a typical value of nside=64, the sky is divided into 49,152 pixels. The pixelation of the line-of-sight (LOS) redshift prior reduces the bias caused by variations in galaxy catalogue incompleteness -- the completeness is different for sky areas explored by different galaxy surveys that make up the GLADE galaxy catalogue \cite{GLADE+}. Assuming a uniform completeness across the sky will overestimate or underestimate the completeness of some areas, creating bias in GW host identification. Therefore using galaxy catalogue with specific completeness in each pixel, which is defined according to the apparent magnitude threshold of the survey, will reduce such bias. 

The final form of the posterior with the dark siren approach is given by 
\begin{equation}\label{Eq:mcmc_framework}
\begin{aligned}
p(\Lambda|\{x_\text{GW}\},\{D_\text{GW}\},I) 
& \propto p(\Lambda|I) p(N_\text{det}|\Lambda,I) \left[\iint p(D_\text{GW}|z,\theta,\Lambda,I) p(\theta|\Lambda,I) \sum^{N_\text{pix}}_j  p(z|\Omega_j,\Lambda,I) d\theta dz \right]^{-N_\text{det}} \\ &\times \prod^{N_\text{det}}_i \left[ \iint \sum^{N_\text{pix}}_j p(x_{\text{GW}i}|z,\Omega_j,\theta,\Lambda,I) p(\theta|\Lambda,I) p(z|\Omega_j,\Lambda,I) d\theta dz \right],
\end{aligned}
\end{equation}
where $p(z|\Omega_j,\Lambda,I)$ is the LOS redshift prior for each pixel at a sky location $\Omega_j$, which can be pre-computed before estimating the posterior, of which the details are presented in \cite{Gray:2023wgj}. In this paper, the GLADE+ galaxy catalogue in the $K$ band is used for the analysis. Although the $K$ band is less complete than the $B_J$ band, its incompleteness model is better understood than the $B_J$ band, and hence gives more reliable results. Unlike the $K$ band, the $B_J$ band suffers from an issue that its incompleteness is not well-captured by a simple magnitude threshold. 

In the GR version of \gwcosmo, $H_0$ is the only cosmological parameter in $\Lambda$. Here we extend \gwcosmo to the estimation of sets of modified gravity parameters such as $\Xi_0$, $n$ in Eq. \eqref{eq:dgw_Xi}, $c_M$ in Eq. \eqref{eq:dgw_cM}, or $D$ in Eq. \eqref{eq:dgw_D}. Since these parameters modify $d_L^{\rm GW}(z)$, but leave $d_L^{\rm EM}(z)$ unchanged, the LOS redshift prior from the galaxy catalogue is not affected. However, modification is required when marginalising over the source frame masses of compact object binaries, because $z$ inferred from $d_L^{\rm GW}(z)$ now differs from that from $d_L^{\rm EM}(z)$ under different values of modified gravity parameters. Inside a pixel, the numerator integrand of the likelihood is given by
\begin{equation}
    p(x_{\text{GW}i}|z,\Lambda,I) = \int_{m_{\rm min}}^{m_{\rm max}} \int_{m_{\rm min}}^{m_1^{\rm s}} p(x_{\text{GW}i}|z,m_1^{\rm s},m_2^{\rm s},\Lambda,I) p(m_1^{\rm s},m_2^{\rm s}|\Lambda,I)dm_2^{\rm s}dm_1^{\rm s},
\end{equation}
where $m_1^{\rm s}$ and $m_2^{\rm s}$ are the component masses in the GW source frame, with the assumption $m_1^{\rm s}>m_2^{\rm s}$. $m_{\rm min}$ and $m_{\rm max}$ are the minimum and the maximum mass for a source of that type, given by the population model. The prior on source masses $p(m_1^{\rm s},m_2^{\rm s}|I)$ is inferred from the compact object mass distribution models \cite{Abbott_2021_population}. However, the posterior samples from GW observation data are generated in the detector frame. In addition, it is $d_L^{\rm GW}(z)$ that is measured from the GW signal instead of $z$. Therefore we have
\begin{equation}
    p(x_{\text{GW}i}|z,\Lambda,I) = \int_{m_{\rm min}}^{m_{\rm max}} \int_{m_{\rm min}}^{m_1^{\rm s}} p(x_{\text{GW}i}|d_L^{\rm GW}(z,\Lambda),m_1^{\rm det}[m_1^{\rm s},z],m_2^{\rm det}[m_2^{\rm s},z],I) p(m_1^{\rm s},m_2^{\rm s}|I)dm_2^{\rm s}dm_1^{\rm s},
    \label{eq:px}
\end{equation}
Applying Bayes' theorem we obtain
\begin{equation}
    p(x_{\text{GW}i}|d_L^{\rm GW},m_1^{\rm det},m_2^{\rm det},I) = \frac{p(d_L^{\rm GW},m_1^{\rm det},m_2^{\rm det}|x_{\text{GW}i},I)p(x_{\text{GW}i}|I)}{\pi(d_L^{\rm GW},m_1^{\rm det},m_2^{\rm det}|I)} \propto \frac{p(d_L^{\rm GW},m_1^{\rm det},m_2^{\rm det}|x_{\text{GW}i},I)}{\pi(d_L^{\rm GW},m_1^{\rm det},m_2^{\rm det}|I)},
    \label{eq:pdgw}
\end{equation}
where $\pi(d_L^{\rm GW},m_1^{\rm det},m_2^{\rm det}|I)$ is the prior applied to $d_L^{\rm GW},m_1^{\rm det}$ and $m_2^{\rm det}$ when computing the posterior samples of a GW event. The posterior samples are effectively the numerator of Eq. \eqref{eq:pdgw}. For the GWTC-3 data set, the posterior samples are generated with two options: $\pi(d_L^{\rm GW})\propto(d_L^{\rm GW})^2$ or $\pi(d_L^{\rm GW})$ uniform in comoving volume prior. In this paper we used the data set with $\pi(d_L^{\rm GW})\propto(d_L^{\rm GW})^2$. Meanwhile $\pi(m_1^{\rm det},m_2^{\rm det})$ is uniform. Then we convert the posterior samples of $d_L^{\rm GW}$ into those of $z$ by
\begin{equation}
    p(d_L^{\rm GW}(z,\Lambda),m_1^{\rm det},m_2^{\rm det}|x_{\text{GW}i},I) = p(z,m_1^{\rm det},m_2^{\rm det}|x_{\text{GW}i},\Lambda,I) \bigg|\frac{\partial z}{\partial d_L^{\rm GW}} \bigg|.
\end{equation}
The conversion from $d_L^{\rm GW}$ to $z$, and the Jacobian $|{\partial z}/{\partial d_L^{\rm GW}}|$ for posterior samples needs to be modified under different parameterizations, which is the main modification compared to the GR version of \gwcosmo. Finally, the transformation from source masses into detected masses gives
\begin{equation}
    dm_1^{\rm s} dm_2^{\rm s} = \bigg| \frac{\partial(m_1^{\rm s}, m_2^{\rm s})}{\partial(m_1^{\rm det}, m_2^{\rm det})} \bigg| dm_1^{\rm det} dm_2^{\rm det} = \frac{1}{(1+z)^2}dm_1^{\rm det} dm_2^{\rm det}
\end{equation}
Following the steps above, the expression for the integrand in Eq. \eqref{eq:px} becomes (remember this is part of the numerator in Eq. \eqref{Eq:mcmc_framework}):
\begin{equation}
\begin{aligned}
    p(x_{\text{GW}i}|z,\Lambda,I) \propto \int_{m^{\rm det}(m_{\rm min},z)}^{m^{\rm det}(m_{\rm max},z)} \int_{m^{\rm det}(m_{\rm min},z)}^{m^{\rm det}(m_1^s,z)} & \frac{p(z,m_1^{\rm det},m_2^{\rm det}|x_{\text{GW}i},\Lambda,I)}{(d_L^{\rm GW})^2} \bigg|\frac{\partial z}{\partial d_L^{\rm GW}} \bigg| \\
    & \times \frac{p(m_1^{\rm s}[m_1^{\rm det},z],m_2^{\rm s}[m_2^{\rm det},z]|I)}{(1+z)^2} dm_2^{\rm det}dm_1^{\rm det}.
    \label{eq:num_int}
\end{aligned}
\end{equation}
The posterior samples on $z$ computed from $d_L^{\rm GW}$ given $\Lambda$, $m_1^{\rm det}$ and $m_2^{\rm det}$ for a GW event can be approximated as a sum
over delta-functions:
\begin{equation}
    p(z,m_1^{\rm det},m_2^{\rm det}|x_{\text{GW}i},\Lambda,I) \approx \frac{1}{N_{\rm pos}} \sum^{N_{\rm pos}}_{k=1} \delta(z-z_k(d_{L,k}^{\rm GW},\Lambda))\delta(m_1^{\rm det}-m_{1,k}^{\rm det})\delta(m_2^{\rm det}-m_{2,k}^{\rm det}),
\end{equation}
where $k$ denotes the $k^{\rm th}$ posterior sample and $N_{\rm pos}$ denotes the total number of posterior samples. Inserting this back to Eq. \eqref{eq:num_int} we obtain the result as
\begin{equation}
    p(x_{\text{GW}i}|z,\Lambda,I) \propto \frac{1}{N_{\rm pos}} \sum^{N_{\rm pos}}_{k=1} p(m_1^{\rm s}(m_{1,k}^{\rm det},z),m_2^{\rm s}(m_{2,k}^{\rm det},z)|I) \bigg|\frac{\partial z}{\partial d_L^{\rm GW}} \bigg| \frac{1}{(d_L^{\rm GW})^2(1+z)^2}\delta(z-z_k(d_{L,k}^{\rm GW},\Lambda)).
\end{equation}
 The above quantity is then multiplied by the LOS redshift prior\footnote{In reality, a kernel density estimate (KDE) of $p(x_{\text{GW}i}|z,\Lambda,I)$ is created first, and multiplied with the redshift prior.} $p(z|\Omega_j,\Lambda,I)$, and integrated over redshift. Summing the results over all pixels one obtains the numerator of the likelihood as indicated in Eq. \eqref{Eq:mcmc_framework}. The final result of the numerator is given by
 \begin{align}
    p(x_{\text{GW}i}|\Lambda,I) \propto \frac{1}{N_{\rm pos}} \sum^{N_{\rm pos}}_{k=1} p(m_1^{\rm s}(m_{1,k}^{\rm det},z_k),m_2^{\rm s}(m_{2,k}^{\rm det},z_k)|I) \bigg|\frac{\partial z}{\partial d_L^{\rm GW}} &\bigg|_{z=z_k} \frac{1}{(d_{L,k}^{\rm GW})^2(1+z_k)^2} \nonumber\\
    & \times\sum^{N_\text{pix}}_j  p(z_k|\Omega_j,\Lambda,I).
\end{align}
The details of computing the LOS redshift prior $p(z_k|\Omega_j,\Lambda,I)$ are described in \cite{Gray:2023wgj}. Similarly, injections in the detector frame need to be converted into the source frame when marginalising the detection probability in the denominator of the likelihood over redshift and component masses. The detection probability is given in the same form by
 \begin{align}
    p(D_{\text{GW}}|\Lambda,I) \propto \frac{1}{N_{\rm inj}} \sum^{N_{\rm inj}}_{k=1} p(m_1^{\rm s}(m_{1,{\rm inj},k}^{\rm det},z_k),m_2^{\rm s}(m_{2,{\rm inj},k}^{\rm det},z_k)|I) &\bigg|\frac{\partial z}{\partial d_L^{\rm GW}} \bigg|_{z=z_{{\rm inj},k}} \frac{1}{(d_{L,{\rm inj},k}^{\rm GW})^2(1+z_{{\rm inj},k})^2} \nonumber\\
    & \times\sum^{N_\text{pix}}_j  p(z_{{\rm inj},k}|\Omega_j,\Lambda,I).
 \end{align}
By inserting the results of the numerator and the denominator back into Eq. \eqref{Eq:mcmc_framework}, and multiplying over all the GW events, we can thus obtain the posterior on the parameter set $\Lambda$.

\subsection{Prior on source masses and merger rate evolution \label{sec:prior_mass_rate}}

\noindent The source mass prior $p(m_1^{\rm s},m_2^{\rm s}|I)$ is obtained from the mass distribution model. There are several different BBH mass distribution models adopted in the LVK analysis, such as the Broken Power Law model and the Power Law + Peak model \cite{GWTC-3:cosmology,Abbott_2021_population}. In our analysis, we use the Power Law + Peak model for black hole distribution, because it is preferred over other models from the population analysis of O3 data \cite{KAGRA:2021duu}, and is adopted by a number of similar works \cite{Leyde_2022,mastrogiovanni2023novel,Gray:2023wgj}. We use a uniform distribution for neutron stars as in the LVK GWTC-3 cosmology paper \cite{GWTC-3:cosmology}. There are 8 parameters in the description of the Power Law + Peak model. In our analysis of GWTC-3 events in Section \ref{sec:GWTC-3}, we will study cases where these parameters are fixed or varied. In the cases where they are varied, the priors on them are uniformly distributed, and are listed in Table \ref{tab:prior} in Section \ref{sec:GWTC-3}.

The explicit redshift prior actually depends on the presence of GW sources, which we denote by $s$. By Bayes' theorem it is given that
\begin{equation}
    p(z|s,\Omega_j,\Lambda,I) = \frac{p(s|z,\Omega_j,\Lambda,I)p(z|\Omega_j,\Lambda,I)}{p(s|\Omega_j,\Lambda,I)}.
\end{equation}
The term $p(s|\Omega_j,\Lambda,I)$ is cancelled for appearing in both the numerator and the denominator of Eq. \eqref{Eq:mcmc_framework}. Assuming that GW sources are distributed uniformly in the sky on the scale of our pixels, the dependence on $\Omega_j$ can be dropped. Then $p(s|z)$ depends on the redshift evolution of the merger rate of compact object binaries. Following the LVK cosmology paper \cite{GWTC-3:cosmology}, we adopt a parameterization motivated by connection of GW events to the star formation rate in \cite{Madau_2014}, which has the form
\begin{equation}
    p(s|z,\gamma,k,z_p) = [1+(1+z_p)^{-\gamma-k}] \frac{(1+z)^\gamma}{1+[(1+z)/(1+z_p)]^{\gamma+k}}.
\end{equation}
The free parameters $\gamma$, $k$ and $z_p$ control the merger rate redshift evolution, and are included in the hyper-parameters to be measured. The merger rate evolution can be scaled by an overall factor $R_0$, but it is not included in the current \gwcosmo analysis.

\section{Tests on Mock Data \label{sec:mock_data}}

\subsection{Simulation setup}

\noindent Because of the limited number of detected GW events at the present time, we first use mock GW events to test the measurement of modified gravity parameters by \gwcosmo. We use the First Two Years of Electromagnetic Follow-Up with Advanced LIGO and Virgo (F2Y) mock data set \cite{Singer_2014} to test our bright siren methodology. This catalogue was generated by injecting around 50,000 BNS mergers with the LIGO and Virgo sensitivity curves mimicking the O2 run. The criteria for a detected bright siren is having single detector SNR over 4, and a network SNR over 12. As a result only approximately 500 BNS events are marked as detected events, among which 250 of them are randomly selected for parameter estimation. We used these 250 bright sirens for our test of gravity with \gwcosmo; we note this is the same mock data challenge as employed for $H_0$ in \cite{PhysRevD.101.122001}.

We generated our own BNS injections for computing the detection probability. We used the same O2 LIGO and Virgo sensitivity curves as in \cite{Singer_2014}, which are the “early” and “mid” sensitivity curves for L1 and H1 in LIGO technical document \cite{H1_L1_O2}, and the O2 noise curve for V1 from \cite{KAGRA:2013rdx}. Each of the three detectors has a duty factor of 76.8\% as indicated in Fig. 9 of \cite{Singer_2014}. We generated 50,000 injections with the SNR threshold of 12, GW luminosity distance between 0 and 500 Mpc, and the BNS component mass between $1.2$ and $1.6\, M_\odot$, assuming $m_1>m_2$. 

A significant feature of the F2Y dataset is that it is generated with a local universe approximation, given that the redshifts of detectable BNS events are expected to be very low. The linear cosmology approximation $d_L^{\rm EM} \approx cz/H_0$ is applied for the EM counterparts of BNS injections. The BNS $d_L^{\rm GW}$ is then parameterized with the linearized $d_L^{\rm EM}$ in the three modified gravity models introduced in Section \ref{sec:method}. We note that in the case of the Hubble constant analysis, the detection probability under a low redshift approximation simply grows as $H_0^3$ \cite{Chen:2017rfc}; in the case of modified gravity it acquires additional corrections. On the other hand, the detected BNS masses have no redshift correction since this is negligible in the nearby universe. The mass prior we used in the analysis is uniform in $[1.2, 1.6]~M_\odot$, since the F2Y BNSs were generated with component masses between 1.2 and 1.6 $M_\odot$. Such a mass prior is much narrower than the range of $[1.0, 3.0]~M_\odot$ used in later LVK analyses: neutron stars were generally expected to have masses strongly peaked around $1.4~M_\odot$ before much heavier neutron stars were detected in events from the O3 observing run.

\subsection{Results}

\begin{figure}[h]
    \centering
    \includegraphics[width=0.6\textwidth]{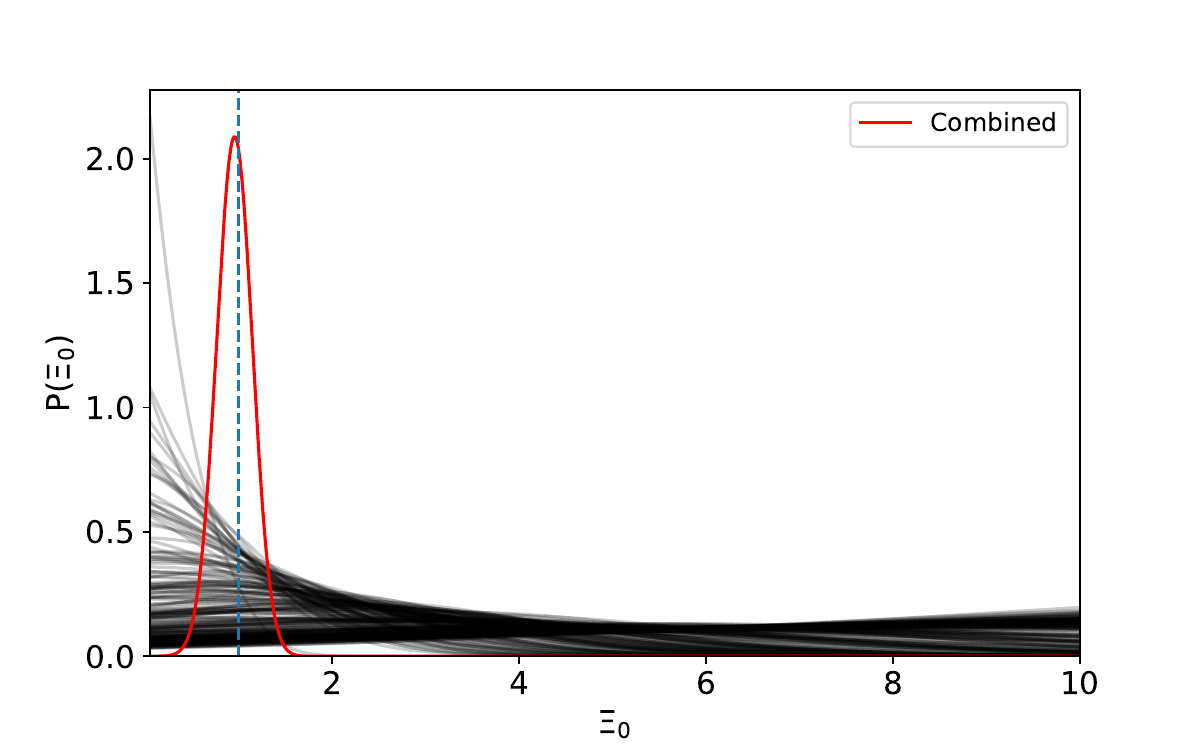}
    \caption{The normalized posterior probability distribution for $\Xi_0$ for individual events (black) and the combined one (red). The blue vertical line labels $\Xi_0=1$ recovery of GR. }
    \label{fig:Xi0_MDC}
\end{figure}

\noindent We first measure $\Xi_0$ in the $(\Xi_0,n)$ parameterization shown in Eq. \eqref{eq:dgw_Xi} with a uniform prior in $[0.05,10]$, while fixing other cosmological parameters as $H_0=70$ ${\rm km}~{\rm s}^{-1}~{\rm Mpc}^{-1}$ and $n=1.91$\footnote{Note that other cosmological parameters such as $\Omega_{M0}$ will not be varied in any of our analyses: they are well-measured by EM probes already, so we adopt their central values.}. The normalized posterior of $\Xi_0$ for each bright siren is plotted in black, and the combined posterior is plotted in red in Fig. \ref{fig:Xi0_MDC}. The measured value with $1\sigma$ bound is $\Xi_0 = 0.96^{+0.19}_{-0.19}$, which is consistent with GR. Our results show that \gwcosmo is capable of constraining $\Xi_0$ well in the simplified case of a one-parameter analysis with a number of bright sirens. The number of bright sirens needed to obtain this bound (250) may seem quite large; this is because, as discussed in section \ref{subsec:Xi0n}, $\Xi_0$ is most relevant to high-redshift GW events, and the mock bright sirens used here are all at low redshifts.

However, the 1D measurement of $\Xi_0$ can be misleading, because it has assumed $H_0$ to be a fixed value. 
Since $H_0$ and $\Xi_0$ both determine the conversion from GW distances to redshifts, they are expected to be degenerate. Ideally it is more sensible to perform a 2D $H_0$-$\Xi_0$ joint analysis. However, the number of posterior samples for the F2Y mock BNS is $\sim 1000$, which is very low compared to the real events that have tens of thousands of posterior samples. Therefore the estimation of redshift for F2Y data is less accurate. Our joint analysis shows that the GR value $(H_0=70,\Xi_0=1)$ is on the edge of the $3\sigma$ bound of the 2D joint posterior. Due to the low number of posterior samples for the F2Y data, the 2D result is unreliable.

\begin{figure}[h]
    \begin{subfigure}[b]{0.48\textwidth}
    \centering
    \includegraphics[width=\textwidth]{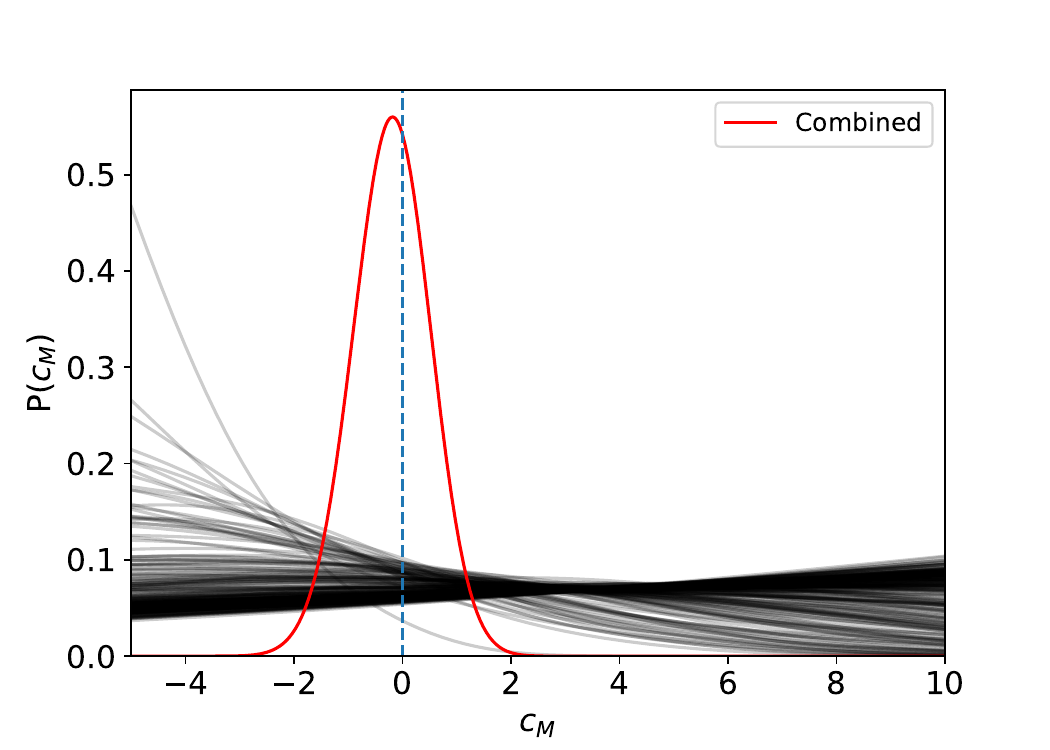}
    \end{subfigure}
    \begin{subfigure}[b]{0.48\textwidth}
    \centering
    \includegraphics[width=\textwidth]{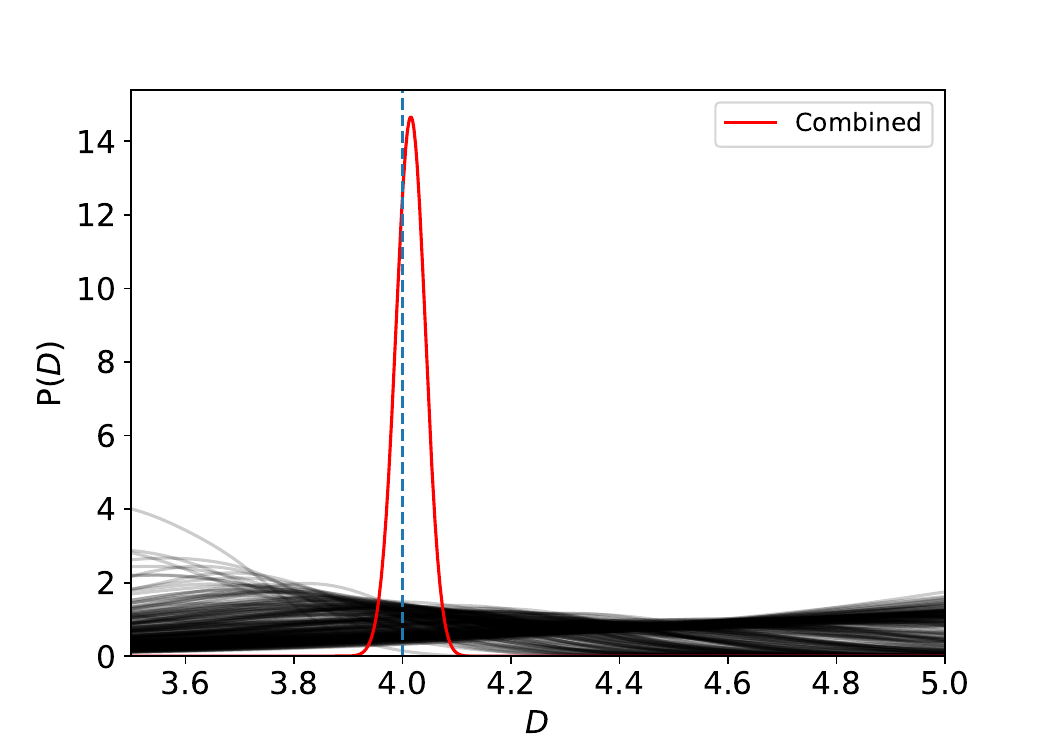}
    \end{subfigure}
    \caption{The normalized posterior probability distribution for $c_M$ (left panel) and $D$ (right panel) for individual events (dark) and the combined constraint (red). The blue vertical lines indicate the GR limits of $c_M=0$ and $D=4$. }
    \label{fig:cM_D_MDC}
\end{figure}

We further perform the same analysis on $c_M$ in the Horndeski class parameterization shown in Eq.\eqref{eq:dgw_cM} and $D$ in the extra-dimension parameterization shown in Eq.\eqref{eq:dgw_D}. We fix $H_0$, mass distribution and merger rate redshift evolution parameters as in the $\Xi_0$ measurement. In the measurement of $c_M$, the values of $\Omega_{m,0}$ and $\Omega_{\Lambda,0}$ are needed to reconstruct the evolution of the fraction of dark energy density as described in Eq. \eqref{eq:alphaM_cM}. We adopt $\Omega_{m,0}=0.3065$ from the Planck result \cite{Planck2016} as in the LVK cosmology paper \cite{GWTC-3:cosmology}, and $\Omega_{\Lambda,0}=1-\Omega_{m,0}$ as other components of energy density are negligible today. We apply the uniform prior for $c_M$ in [-5, 10]. The normalized posterior of $c_M$ for individual events and the combined posterior are plotted in the left panel of Fig. \ref{fig:cM_D_MDC}. The $1\sigma$ bound measured value is $c_M=-0.18^{+0.71}_{-0.72}$, which is consistent with GR where $c_M=0$.

In the measurement of $D$ in the extra-dimension parameterization shown in Eq. \eqref{eq:dgw_D}, we choose to fix $R_c=100$ Mpc, which is the same magnitude as that of the luminosity distances for BNS events, so that the modified gravity effect can be probed at such distances. This value may not match the one motivated from fundamental physics, but it allows us to test the constraint on $D$ at a scale of the BNS mock data we have available. In addition we also fix $n_D=1$. The prior of $D$ is uniform in [3.5, 5.0]. The effect of varying $D$ with fixed $R_c$ and $n_D$ is shown in Fig. \ref{fig:dgw_z}. The normalized posterior of $D$ for individual events and the combined posterior are plotted in the right panel of Fig. \ref{fig:cM_D_MDC}. The $1\sigma$ bound measured value is $D=4.015^{+0.027}_{-0.027}$, which is consistent with the GR limit of this class of theories.

In summary, we obtain good 1D constraints on modified gravity parameters in all of the three alternative gravity models using 250 mock bright sirens, and recover consistency with GR in all cases (as of course expected, as the mock data is generated under GR). However, because of the low number of posterior samples of the F2Y data, 2D joint constraints of $H_0$ and modified gravity parameters are not reliable. Still, our test shows that \gwcosmo is capable of measuring deviations beyond GR with the bright siren method. It will be used to test gravity with bright sirens that are expected to be detected in the O4 run.

\section{Reanalysing the GWTC-3 data\label{sec:GWTC-3}}

\subsection{Parameter ranges}

\noindent In this section we show the results of reanalysing the GWTC-3 events with \gwcosmo, including modified gravity parameters. We use the same 46 dark sirens from the GWTC-3 catalogues as in the LVK GWTC-3 cosmology paper \cite{GWTC-3:cosmology} with net SNR larger than 11, including 42 BBH mergers, 1 BNS merger GW190425, and 3 NSBH mergers GW190814, GW200105\_162426 and GW200115\_042309. The posterior samples of GWTC-2.1 \cite{GWTC-2.1} are used in computing the numerator of the likelihood for events from O1 to O3a, and GWTC-3 samples \cite{GWTC-3} for events in O3b. 
The denominator of the likelihood is computed using $2\times 10^6$ GW injections with the SNR threshold set to 11. The injected detected masses range from 1 to 500 $M_\odot$, and the injected GW distances are varied between 0.1 and $\sim 20,000$ Mpc. Such a high value for the maximum GW distance is needed to cover the GW distance for extreme values of the modified gravity parameters for $z\in (0,10)$.

The LOS redshift prior we used is pre-computed using the GLADE+ galaxy catalogue in the $K$ band, with resolution of nside = 64. We will focus on the $\Xi_0-n$ parameterization for the next few subsections, but the other modified gravity parameters will be treated analogously, and their final results will be discussed in Section \ref{sec:combine_BBH_NSBH}.
Given the maximum redshift $z=10$ of the LOS redshift prior, we need to choose joint prior bounds of $H_0$ and $\Xi_0$ such that the maximum injected GW distance corresponds to redshift not larger than 10. As a result, we vary $\Xi_0$ in a range of [0.3,10] when fixing $H_0=70~{\rm km}~{\rm s}^{-1}~{\rm Mpc}^{-1}$. A combination of the maximum $H_0$ and the minimum $\Xi_0$ gives the largest redshift for the maximum injected GW distance, so we need to reduce the higher bound of $H_0$ or raise the lower bound of $\Xi_0$ when varying both of them. We chose a range of [20,140] ${\rm km}~{\rm s}^{-1}~{\rm Mpc}^{-1}$ for $H_0$ and [0.35,10] for $\Xi_0$ for the joint analysis.

\subsection{Dark sirens results}

\subsubsection{1D measurement of $\Xi_0$}

\begin{figure}
    \centering
    \includegraphics[width=\textwidth]{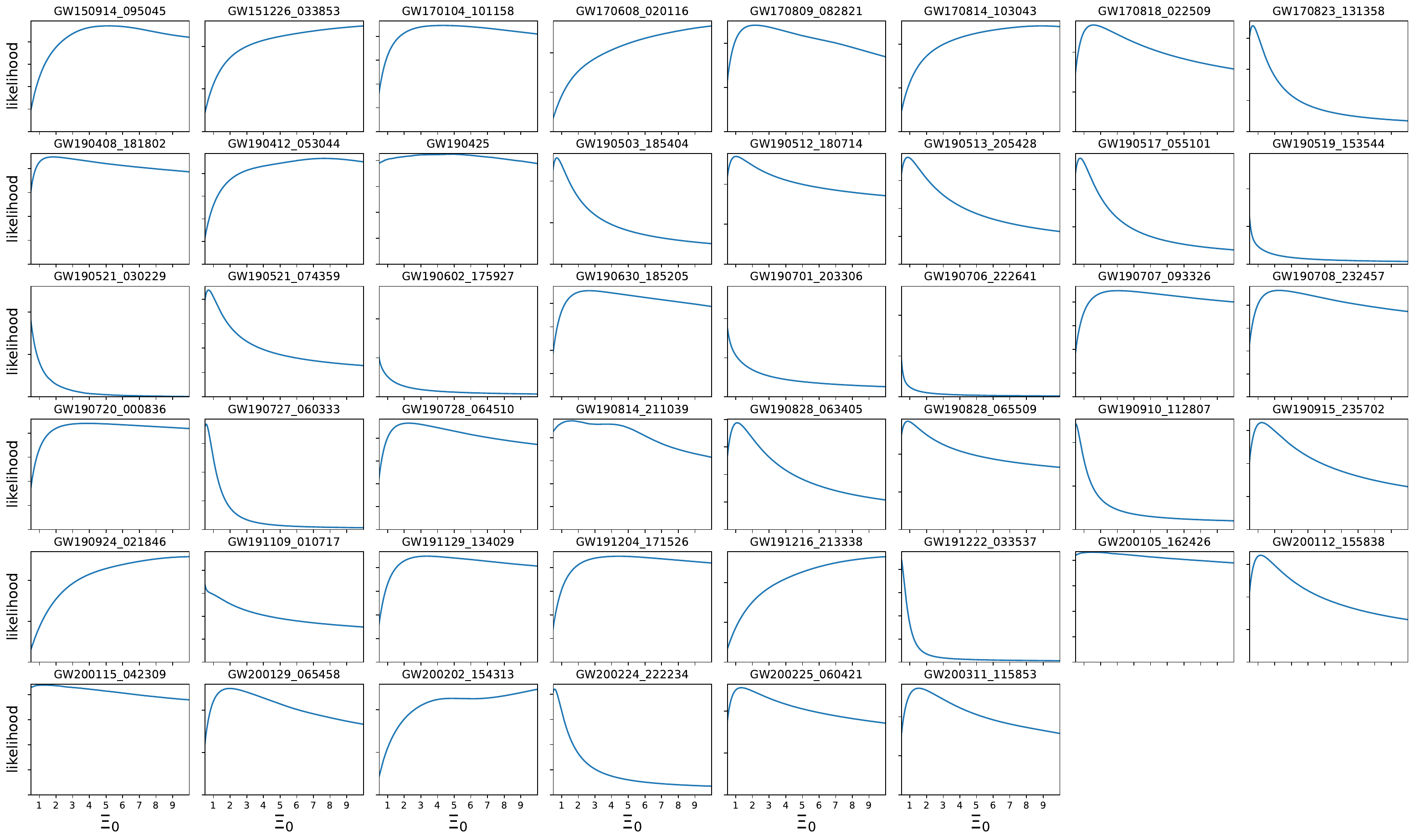}
    \caption{The one-dimensional likelihood of $\Xi_0$ from individual selected events used in our analysis. Note that all other parameters are held fixed. These 1D curves demonstrate the range of possible curve shapes obtained when galaxy catalogue support is moderately weak; see Appendix \ref{sec:posterior_samples} for a detailed explanation.}
    \label{fig:Xi0_individual}
\end{figure}

\noindent We first measure the posterior of $\Xi_0$ in Eq. \eqref{eq:dgw_Xi}
with a flat prior on [0.3,10], while fixing the other two cosmological parameters to be $H_0=70~{\rm km}~{\rm s}^{-1}{\rm Mpc}^{-1}$ and $n=1.91$. We also fix the parameters in the Power Law + Peak mass prior as in the LVK cosmology paper \cite{GWTC-3:cosmology}, which are $\alpha=3.78$, $\beta=0.81$, $m_{\rm min}^{\rm BH}=4.98M_\odot$, $m_{\rm max}^{\rm BH}=112.5M_\odot$, $\delta_m=4.8M_\odot$, $\mu_g=32.27M_\odot$, $\sigma_g=3.88M_\odot$ and $\lambda_p=0.03$. A brief description of the physical significance of these parameters is given in Table~\ref{tab:prior}. In the cases of NSBH and BNS, we fix $m_{\rm min}^{\rm NS}=1.0M_\odot$ and $m_{\rm max}^{\rm NS}=3.0M_\odot$. The merger rate evolution parameters are also fixed with the same values $\gamma=4.59$, $k=2.86$ and $z_p=2.47$. The normalized posterior of $\Xi_0$ for each individual event is plotted in Fig.~\ref{fig:Xi0_individual}. We can see that there are three typical patterns in the distribution of $p(\Xi_0)$ for different events: favoring low $\Xi_0$ such as GW190910\_112807, favoring high $\Xi_0$ such as GW191216\_213338, and peaking at a certain value of $\Xi_0$ such as GW170809\_082821. The reasons for these patterns can be found by looking at the posterior sample distribution in $z$ converted from $d_L^{\rm GW}$ for different $\Xi_0$ for each event, which we will elaborate on in Appendix \ref{sec:posterior_samples}. The measured value from the combined posterior with $1\sigma$ bound is $\Xi_0=1.06^{+0.25}_{-0.18}$, which is consistent with GR.

\begin{figure}
    \centering
    \includegraphics[width=0.5\textwidth]{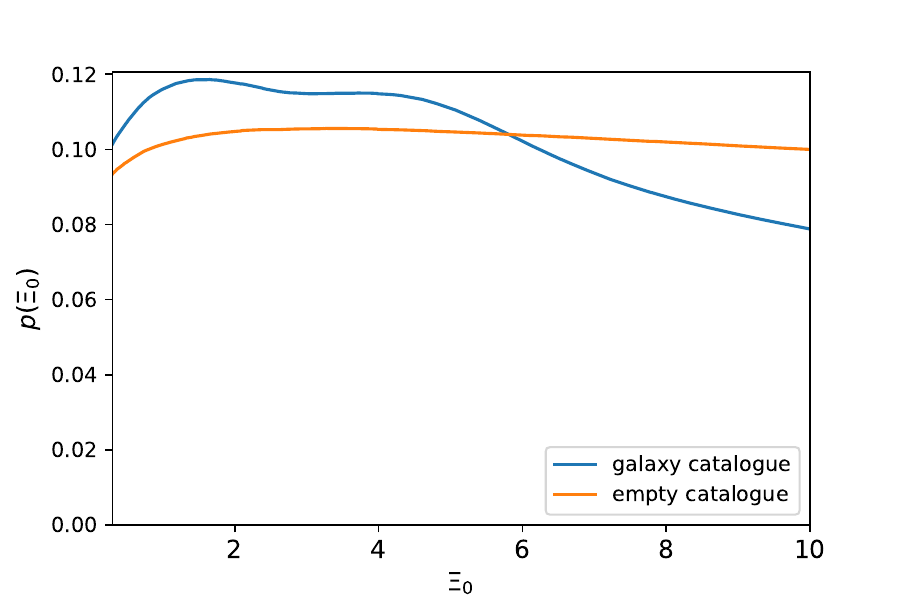}
    \caption{The 1D posterior of $\Xi_0$ for GW190814, with both the GLADE+ galaxy catalogue and an empty catalogue analysis (not galaxies. We see that for this event the clustering in the galaxy catalogue (dis)favours certain values of $\Xi_0$, creating a bumpy structure in the likelihood. Comparable features can be seen in the $H_0$ likelihood shown in Fig.~9 of \cite{Gray:2023wgj}.}
    \label{fig:Xi0_190814}
\end{figure}

For most dark siren events in GWTC-3, the redshift support from the galaxy catalogue does not have a significant effect on improving the precision of the $\Xi_0$ measurement. The event with the most galaxy support is GW190814, which is shown in Fig. \ref{fig:Xi0_190814}; here we compare the analysis with the galaxy catalogue and an empty catalogue (no galaxies). $p(\Xi_0)$ with galaxy catalogue support peaks at lower $\Xi_0$ compared to the flat distribution without the support, which shows that galaxy catalogues have the potential to increase the precision of modified gravity constraints with dark sirens. The galaxy catalogue information has only a weak effect on the current dark siren analysis; this is because the present localization of GW events has a relatively 
low accuracy, so that the information from clustering of galaxies gets washed out as it is averaged over too many pixels. If the localization is small enough, our method is sensitive to clustering of galaxies so that the results will favour some values of $H_0$ and $\Xi_0$.
In addition, the galaxy catalogue is incomplete, especially at high redshift, so it provides limited support to GW events further away.

In future, with an increase in the sensitivity of the KAGRA detector, and the addition of LIGO India to the terrestial network, the localization of GW events will be improved. Moreover, galaxy surveys by new detectors such as the Dark Energy Spectroscopic Instrument (DESI) \cite{Dey_2019} and the Euclid space telescope \cite{Euclid} in the future will provide millions of further spectroscopic galaxy redshifts, particuarly at higher redshifts where catalogue support is currently lacking. This will further accelerate the precision of dark sirens constraints.

\subsubsection{BBH joint analysis}

\begin{figure}[h]
    \centering
    \includegraphics[width=\textwidth]{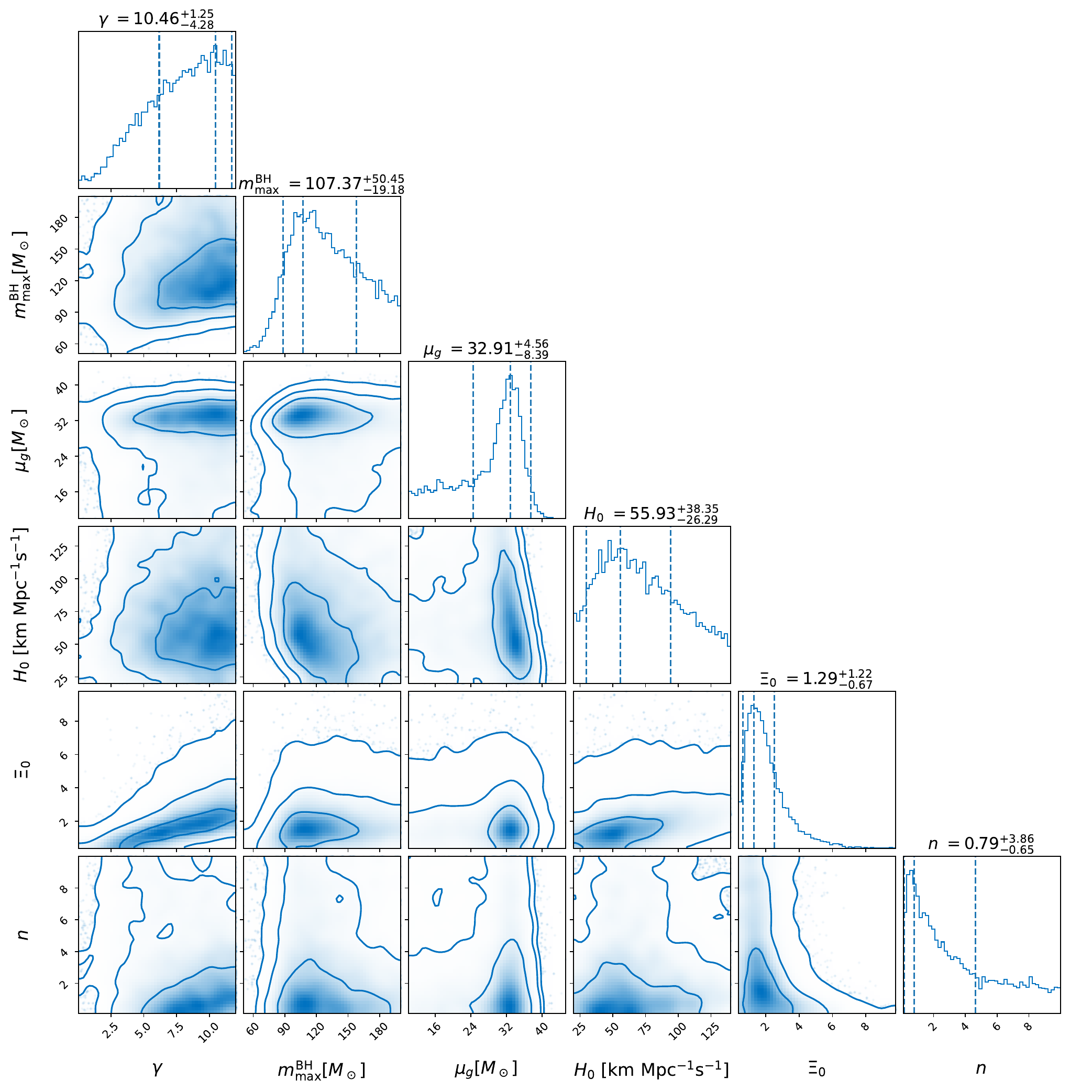}
    \caption{Selected corner plot of joint posteriors for 42 BBH events from the GWTC-3 catalogue computed by the \gwcosmo sampling run. The contours show 1, 2 and 3$\sigma$ bound for the 2D posteriors. The vertical lines in the 1D histogram plot correspond to the peak values and 1$\sigma$ bound of the parameters in the title. The full corner plot including all parameters is shown in Fig. \ref{fig:full_corner}. A brief description of the mass and merger rate hyper-parameters is given in Table~\ref{tab:prior}.}
  \label{fig:selected_corner}
\end{figure}

Next we perform a joint posterior measurement for \textit{all} hyper-parameters under the $(\Xi_0,n)$ MG parameterization with nested sampling in \gwcosmo. We adopt uniform priors for all hyper-parameters including the mass distribution model and the merger rate redshift evolution model as listed in Table \ref{tab:prior}. We present the corner plot of the joint posteriors for 42 BBH events from nested samplings for 6 of the most interesting parameters in Fig. \ref{fig:selected_corner}, which are $\gamma$, $m_{\rm max}^{\rm BH}$, $\mu_g$, $H_0$, $\Xi_0$ and $n$. The full corner plot with all parameters is shown in Fig. \ref{fig:full_corner} in Appendix \ref{sec:full_corner}. 
We can see that some BBH mass distribution parameters are moderately well-constrained; However, $k$ and $z_p$ in the merger rate redshift evolution model are poorly constrained, which is similar to results in the LVK cosmology paper \cite{GWTC-3:cosmology}. In addition, the estimated value of $H_0$ is $55.93^{+38.35}_{-26.29}~{\rm km}~{\rm s}^{-1}{\rm Mpc}^{-1}$, again an uncertainty similar to previous results. The estimated value for $\Xi_0$ is $1.29^{+1.22}_{-0.67}$, which is consistent with GR; as expected, the result is less constrained than that in the 1D measurement because of co-variance with other parameters. Similar results have been obtained using the \icarogw software \cite{Leyde_2022,mastrogiovanni2023novel}. Meanwhile $n$ is measured to be $0.79^{+3.86}_{-0.65}$, which is lower than the fiducial value of 1.91, but with quite broad uncertainties. The reason that $p(n)$ favors low values of $n$ is likely that the $\Xi(z)$ parameterization reduces to GR when $n=0$, so a shift towards low values of $n$ is an alternative fit to the data (instead of $\Xi_0=1$), if the data are consistent with GR.
Moreover, as shown by the 2D $\Xi_0$-$n$ posterior contour, $\Xi_0$ is better constrained when $n$ is larger. This is because the redshift-dependent term with the inverse power of $n$ in Eq. \eqref{eq:dgw_Xi} becomes less significant when $n$ is large, leaving the constant $\Xi_0$ as the more dominant term. Therefore the constraint is put more directly on $\Xi_0$ when $n$ is large. 

\begin{table}[h]
    \centering
    \begin{tabular}{l|l|l}
        \hline
        Parameter & Definition & Prior \\
        \hline
        $H_0$  & Hubble constant  & ${\cal U}(20.0, 140.0)$ \\
        $\Xi_0$  & Modified gravity parameter controlling high-z limit of  & ${\cal U}(0.35, 10.0)$ \\
         & distance ratio in Eq. \eqref{eq:dgw_Xi} & \\
        $n$  & Modified gravity parameter controlling steepness of & ${\cal U}(0.1, 10.0)$ \\
         & distance ratio in Eq. \eqref{eq:dgw_Xi} & \\
        \hline
       $\alpha$  & The power of the power law component in the primary & ${\cal U}(1.5, 8.0)$ \\
        & mass distribution  & \\
       $\beta$  & The power of the power law component in the mass  & ${\cal U}(-4.0, 6.0)$ \\
        & ratio distribution & \\
       $m_{\rm min}^{\rm BH}~[M_\odot]$ & The minimum mass of the mass distribution & ${\cal U}(2.0, 10.0)$ \\
       $m_{\rm max}^{\rm BH}~[M_\odot]$ & The maximum mass of the mass distribution & ${\cal U}(50.0, 200.0)$ \\
       $\lambda_p$ & Fraction of the model in the Gaussian component & ${\cal U}(0.0, 0.5)$ \\
       $\mu_g$ & Mean of the Gaussian component in the primary mass  & ${\cal U}(10.0, 50.0)$ \\
        & distribution & \\
       $\sigma_g$ & Width of the Gaussian component in the primary mass & ${\cal U}(0.1, 20.0)$ \\
        & distribution & \\
       $\delta_m$ & Range of mass tapering at the lower end of the mass  & ${\cal U}(0.0, 15.0)$ \\
       & distribution & \\
       \hline
       $\gamma$  & The power of the power law distribution of the rate  & ${\cal U}(0.0, 12.0)$ \\
        & evolution before redshift $z_p$ & \\
       $k$  & The power of the power law distribution of the rate & ${\cal U}(0.0, 10.0)$ \\
        & evolution after redshift $z_p$ & \\
       $z_p$  & The redshift turning point between two power law  & ${\cal U}(0.0, 10.0)$ \\
        & distributions & \\
       \hline
    \end{tabular}
    \caption{Prior of the parameters in the Power Law + Peak BBH mass distribution.}
    \label{tab:prior}
\end{table}

From the contour plot we can see that $H_0$ and $\Xi_0$ are partially degenerate with each other. This is because $d_L^{\rm GW} \propto \Xi_0/H_0$ as shown in Eq. \eqref{eq:d_EM} and \eqref{eq:dgw_Xi},
so that in the conversion of posterior samples from $d_L^{\rm GW}$ to $z$ in computing the likelihood, varying $H_0$ and $\Xi_0$ while keeping the ratio $\Xi_0/H_0$ unchanged gives the same redshift. So the result that the estimated value of $H_0$ is higher than that in the LVK cosmology paper \cite{GWTC-3:cosmology} is likely caused by $\Xi_0$ favoring a slightly higher value than GR. Similar degeneracy can also be found in Fig. 13 of \cite{Leyde_2022}. 

Apart from $H_0$, we can also see that $\Xi_0$ is strongly degenerate with $\gamma$. In the computation of the likelihood, the LOS redshift prior is multiplied by the merger rate redshift evolution, which is approximately proportional to $(1+z)^\gamma$ when $z\ll z_p$.  
Given that all GW events in GWTC-3 have redshift less than 1, and from star formation arguments $z_p$ is typically considered to have a value around 2, 
the power of $\gamma$ dominates the redshift evolution of the merger rate. This can also explain why $k$, which is the power of the redshift evolution when $z>z_p$, and $z_p$, are poorly constrained with GWTC-3 events. So the numerator of the likelihood is proportional to $\Xi_0$ and $(1+z)^\gamma$, leading to a strong degeneracy between $\Xi_0$ and $\gamma$. A similar feature is also found in Fig. 12 of the \icarogw results \cite{Leyde_2022}, where the distribution of $p(\gamma)$ with the Power Law + Peak BBH population model is similar to ours. Given that the distribution of $p(\gamma)$ favors a higher value, if we allow the prior bound of $\gamma$ to be larger, the distribution of $p(\Xi_0)$ will migrate to higher values because of the degeneracy. The prior ranges for $\gamma$, $k$ and $z_p$ are chosen to be sufficiently wide in consideration of the effect of a possible time delay between the formation and the merger of the binary.

\subsubsection{NSBH joint analysis}

\begin{figure}[h]
    \centering
    \includegraphics[width=\textwidth]{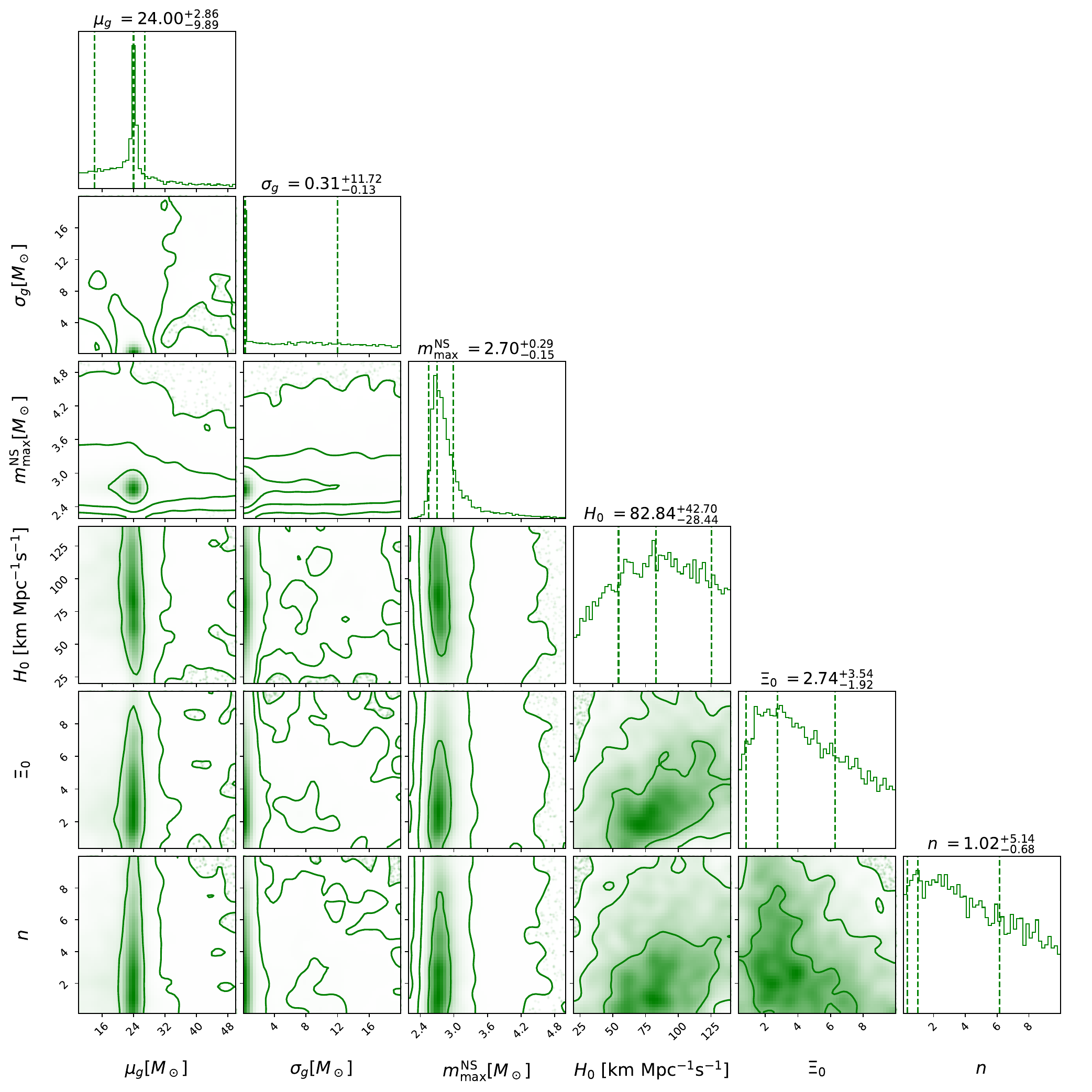}
    \caption{Selected corner plot of joint posteriors for 3 NSBH events from the GWTC-3 catalogue computed by the \gwcosmo sampling run. The contours show 1, 2 and 3$\sigma$ bound for the 2D posteriors. The vertical lines in the 1D histogram plot correspond to the peak values and 1$\sigma$ bound of the parameters in the title. A brief description of the mass and merger rate hyper-parameters is given in Table~\ref{tab:prior}.}
    \label{fig:selected_corner_nsbh}
\end{figure}

We then perform joint analysis on all hyper-parameters with the three NSBH mergers by nested samplings implemented in \gwcosmo. The NSBH events selected are GW190814, GW200105\_162426 and GW200115\_042309. Note that GW190814 is not a confirmed NSBH, but we choose to treat it as one for this analysis. 
In addition to all of the parameters in the analysis with BBHs, parameters $m_{\rm min}^{\rm NS}$ and $m_{\rm max}^{\rm NS}$ for the uniform neutron star population model are included in the estimation. We use the uniform prior of $m_{\rm min}^{\rm NS}\in[0.5,1.5]M_\odot$ and $m_{\rm max}^{\rm NS}\in[1.501,5]M_\odot$. Due to the small number of events, most of the parameters in the BBH population model and the merger rate evolution model are poorly constrained. Furthermore, the marginalized posteriors for some of the parameters show misleading results, as shown in the selected contours in Fig. \ref{fig:selected_corner_nsbh}. For example, the posteriors for $\mu_g$ and $\sigma_g$ indicate a narrow Gaussian peak at $\sim 24M_\odot$ for the black hole population model. This is caused alone by the black hole with mass $\sim 23M_\odot$ in GW190814. The same appearance is probably also shown in the constraint of $m_{\rm max}^{\rm NS}$, which is driven by the assumption that GW190814's secondary mass is a neutron star with mass $\sim 2.6M_\odot$. Meanwhile the constraints on cosmological parameters are generally weaker than those with the BBHs due to the lower number of NSBH events.

\subsection{Combined BBH and NSBH joint analysis \label{sec:combine_BBH_NSBH}}

\begin{figure}[h]
    \centering
    \includegraphics[width=0.6\textwidth]{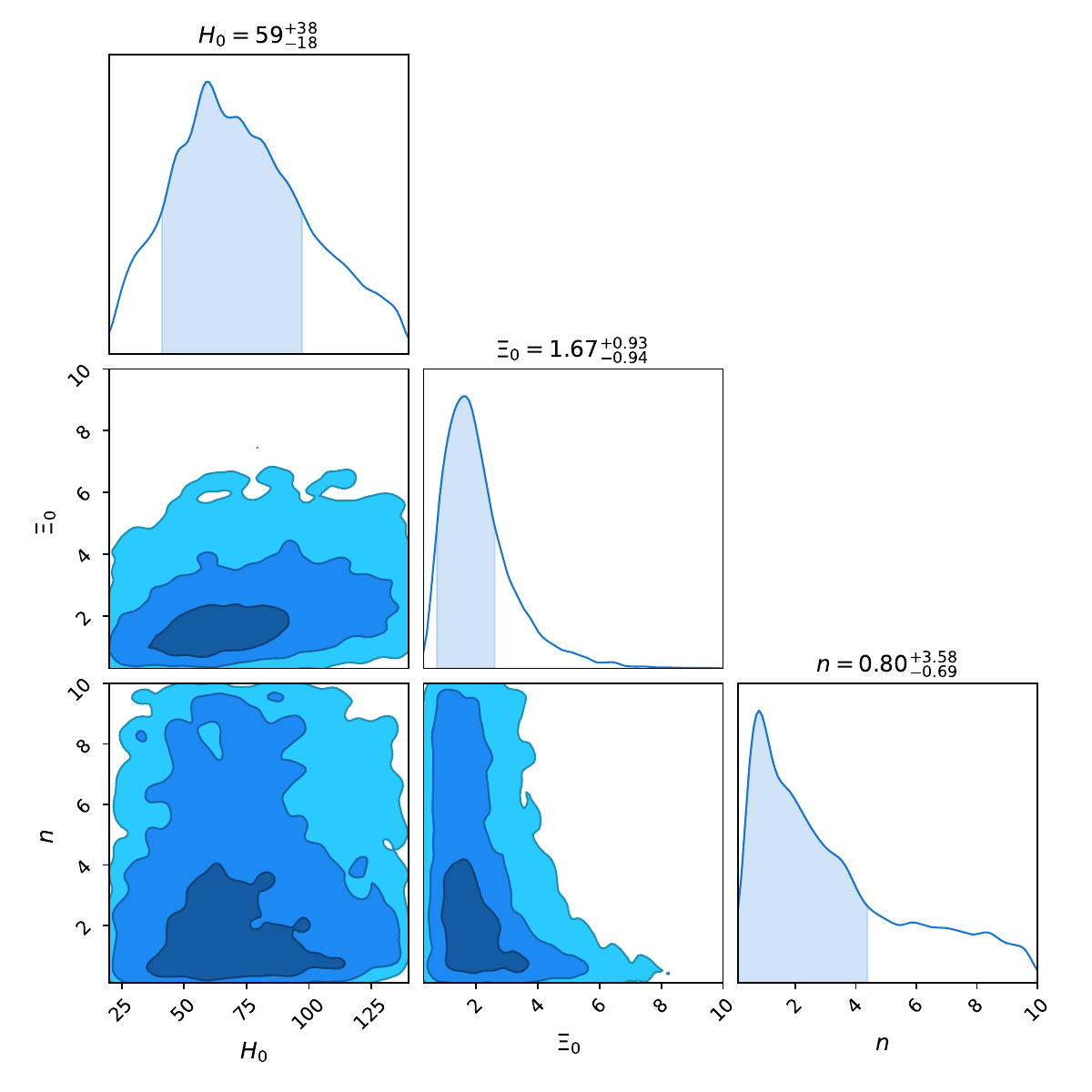}
    \caption{Posteriors of $H_0$, $\Xi_0$ and $n$ by combining results of BBHs and NSBHs from the \gwcosmo sampling run. }
    \label{fig:combined_corner}
\end{figure}

We can improve the constraints on cosmological parameters by combining the results of BBHs and NSBHs. We believe this is the first time a combined dark sirens analysis of BBHs and NSBHs has been applied to tests of gravity. Since the sampling analysis for NSBHs includes two more parameters, $m_{\rm max}^{\rm NS}$ and $m_{\rm min}^{\rm NS}$, the sampling data from BBHs and NSBHs cannot be combined simply. We first compute a kernel density estimation (KDE) function for the multi-dimensional joint posterior of the cosmological parameters with the sampling data. Then we compute the joint posterior for each gridded value of the cosmological parameter space with the KDE for BBHs and NSBHs, and multiply the two posteriors over the space. Finally we obtain the posterior of each parameter by marginalizing the joint multi-dimensional posterior over other parameters. The combined posteriors for $H_0$, $\Xi_0$ and $n$ are shown in Fig. \ref{fig:combined_corner}. The estimated values of $H_0$, $\Xi_0$ and $n$ for the combined posterior are between the BBH posterior and the NSBH posterior, with a slightly smaller $1\sigma$ uncertainty than the individual BBH and NSBH posteriors, which is as expected. Again, we see the partial degeneracies between $\Xi_0$ and $H_0$, and $\Xi_0$ and $n$ are maintained. 

\begin{figure}[h]
    \centering
    \includegraphics[width=0.45\textwidth]{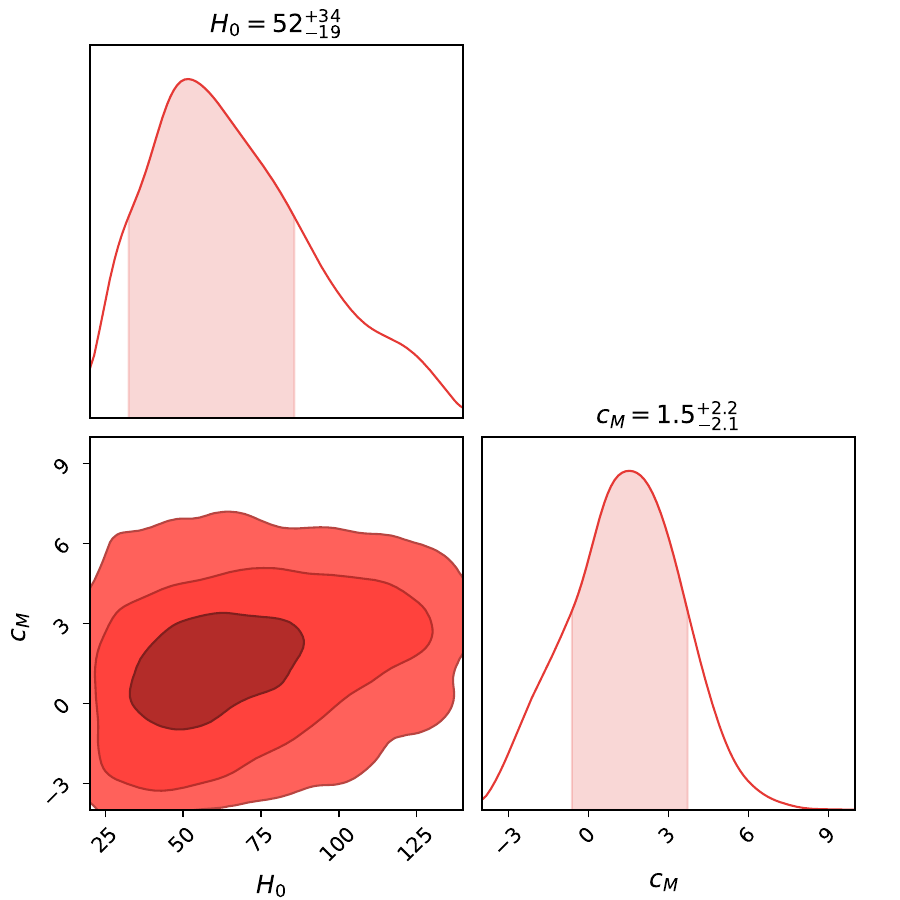}
    \caption{Posteriors of $H_0$ and $c_M$ by combining results of BBHs and NSBHs from the \gwcosmo sampling run. }
    \label{fig:combined_corner_cM}
\end{figure}

\begin{figure}[h]
    \centering
    \includegraphics[width=0.6\textwidth]{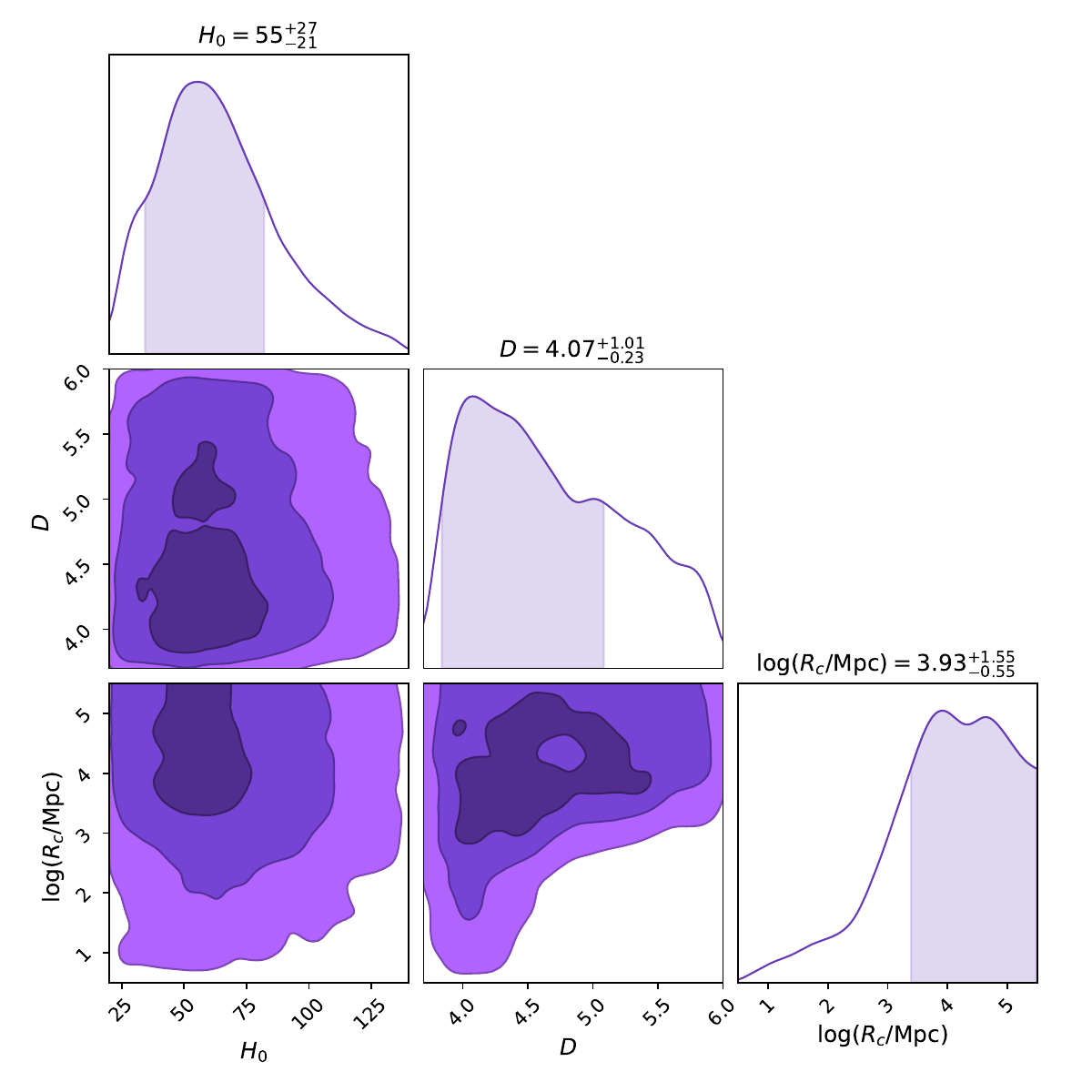}
    \caption{Posteriors of $H_0$, $D$ and $\log R_c$ by combining results of BBHs and NSBHs from the \gwcosmo sampling run. }
    \label{fig:combined_corner_D}
\end{figure}

The results for the other two modified gravity models introduced in this work for the combined posterior of BBHs and NSBHs are shown in Fig. \ref{fig:combined_corner_cM} and  \ref{fig:combined_corner_D}. For the Horndeski class model, we apply a uniform prior for $c_M$ in [-4,10]. The lower prior bound is restricted by the maximum injected GW distance. The combined posterior gives $c_M=1.5_{-2.1}^{+2.2}$, which is consistent with the GR prediction of $c_M=0$ within a $1\sigma$ bound. Principally due to the higher redshifts of the dark siren events, this constraint is much narrower than that obtained by the bright siren GW170817 \cite{Lagos:2019kds}, and is competitive with that from spectral sirens \cite{Ezquiaga:2021ayr}. We can also see slight degeneracy between $H_0$ and $c_M$ in the contour. 

On the other hand, for the extra-dimension model, we use a uniform prior for $D\in [3.7,6]$, $\log (R_c/{\rm Mpc})\in [0.5,6]$ and $n_D\in [0.3,10]$. We obtain $D=4.07_{-0.23}^{+1.01}$, which is consistent with 4-dimensional spacetime within 1$\sigma$ bound, but the distribution of $p(D)$ has a long tail at higher values. This constraint is slightly better than that obtained with the spectral siren method \cite{MaganaHernandez:2021zyc}. Meanwhile the constraint on $R_c$ is $\log (R_c/{\rm Mpc})=3.93_{-0.55}^{+1.55}$, corresponding to $R_c\sim10$ Gpc, which is around the same scale as the Hubble radius. The drop of the posterior of $\log (R_c/{\rm Mpc})$ at the upper bound is artificial due to the over-smoothing of KDE; the distribution of samples in $R_c$ is actually flat at the upper limit, as GR is recovered when $R_c$ approaches infinity. In Fig. \ref{fig:combined_corner_D} we trim the range of $R_c$ where the over-smoothing takes place. We have also included $n_D$ in the joint posterior, but $n_D$ is essentially unconstrained. Compared with the other two modified gravity models, there is little degeneracy among $H_0$, $D$ and $\log R_c$.

\section{Conclusions\label{sec:conclusions}}

\noindent The detection of gravitational waves 
offers us a powerful tool for testing theories of gravity beyond GR. The detection of bright siren GW170817 kicked off these investigations by providing a strong constraint on a particular aspect of modified GW propagation, namely the propagation speed. However, due to its relative proximity, the bounds obtained on other aspects of modified GW propagation remain relatively weak. We stress that an event ideal for constraining the Hubble parameter (e.g. one at low redshift) is not necessarily an ideal system for tests of gravity and fundamental physics. Therefore, high redshift events should not be always discarded in favour of apparently simpler analyses at low redshift.

Although more bright siren detections are expected going forward, their rate and redshift distribution remains currently rather unknown.
The dark sirens technique, which statistically identifies host galaxies of GW events with galaxy catalogues, enables us to avoid pinning our hopes for GW cosmology and tests of fundamental physics on the uncertain bright siren landscape. Although dark sirens are individually less constraining than bright sirens as cosmological probes, their advantage lies in both their number and security (meaning that, they rely only on regular BBH and NSBH mergers, which we are confident will continue to be detected in increasing numbers). Any moderately well-localised GW event can be used for a dark sirens analysis, provided it meets a modest SNR threshold and has satisfactory parameter estimation results. As such, at least tens (or possibly $\sim 100$) of suitable events are expected during the recently-commenced O4 observation run. We can expect to see corresponding improvements in the parameter constraints presented here on the timescale of a year or two.

In this work we described
the modification to the \gwcosmo software pipeline for constraining modified gravity parameters that affect GW luminosity distances, with both bright sirens and dark sirens. A crucial step forward is that these additional MG parameters can now be constrained concurrently with hyper-parameters describing the mass distribution and redshift evolution of compact objects, as well as the Hubble constant. 

To validate our pipeline and explore the constraining power of future data sets, we first performed our tests of gravity on 250 mock bright sirens from the F2Y mock data catalogue. We computed the 1D posteriors on $\Xi_0$, $c_M$ and $D$ in the three modified gravity parameterizations of Section \ref{sec:method} respectively, recovering the expected consistency with GR. The joint posterior on $H_0$ and $\Xi_0$ with the F2Y mock bright sirens remained weak: it is difficult to constrain modified propagation effects with bright sirens located at low redshifts, where changes to the GW luminosity distances are small. 

Turning our attention to the current data, we then computed the posterior probability on the modified gravity parameters in addition to mass distribution and rate evolution hyper-parameters with 46 dark sirens from the GWTC-3 catalogue. The constraint on $\Xi_0$ from the marginalized posterior with 42 BBHs is $1.29_{-0.67}^{+1.22}$, which is consistent with GR. We observed high degeneracy between $\Xi_0$ and $\gamma$ (a parameter controlling the redshift evolution of the merger rate), which makes the choice of the prior bound of $\gamma$ important for tests of gravity. This is a good example of how constraints on cosmology, astrophysics and fundamental physics are entangled, at least for the present.

Finally, we combined for the first time the results of 42 BBHs and 3 NSBHs, which yields $\Xi_0=1.67_{-0.94}^{+0.93}$. Furthermore, the combined posterior of BBHs and NSBHs gives the constraint of $c_M=1.5_{-2.1}^{+2.2}$ in the Horndeski class model, and $D=4.07_{-0.23}^{+1.01}$ in the extra-dimension model, all of which are consistent with GR.

At present, the effects of adding galaxy catalogue information are most pronounced for event GW190814, due to its good localisation and moderately low redshift; they are not highly significant for most other events. This is primarily due to the relatively low redshift range for which the GLADE+ catalogue employed our analysis has high completeness. Therefore, for most of the GW events to date, the bulk of their GW luminosity distance distribution lies beyond the catalogue. The good news is that other spectroscopic galaxy catalogues with higher completeness and redshift range are forthcoming, such as those from the DESI \cite{Dey_2019} and Euclid \cite{Euclid} experiments (see \cite{Palmese:2021mjm} for an initial dark siren analysis with the DESI catalogue). In some sense, this is the much better issue to have: we know that galaxy catalogue completeness will improve dramatically on a timescale of five years or less. We do not currently know what the corresponding improvements for bright siren counts will be. On longer timescales, the measurement of cosmological and modified gravity parameters with direct cross-correlation between GW events and galaxy surveys has been forecast in \cite{Mukherjee:2018ebj,Mukherjee:2019wcg,Mukherjee:2020mha,Mukherjee:2020hyn,Diaz:2021pem} (see also \cite{Canas-Herrera:2021qxs, Fonseca:2023uay}). This can serve as an independent probe in the 3G era.

A secondary effect in the current constraints is localization of GW events, which was particularly large for some events in the first two LVK observing runs. When `averaged' over a large sky area we expect the galaxy clustering signal -- which is key for (dis)favouring some parameter values -- to be washed out. Given the incompleteness issue above, this is not a major stumbling block at present. With developments at the KAGRA and LIGO-India sites progressing, observations by three widely-separated detectors in future will lead to corresponding improvements in event localization, and should prevent this effect from ever becoming a limiting factor.

In this analysis we selected three specific parameterizations of modified GW luminosity distances to study. Whilst these functional forms are commonly used and motivated by general properties of current MG models, they are intended as representative examples only. The analysis pipelines developed in this work can be easily adapted to use alternative parameterizations or specific gravity models, if desired. In future work we intend to broaden the range of options in \gwcosmo, including effects beyond modified GW luminosity distances.

As the fourth LVK observing run progresses, there may well be significant advances or new discoveries in some of the ingredients that make up the dark sirens recipe, e.g. the mass distribution of compact objects, its possible evolution with redshift, or the merger rates of different source types (as analysed with GWTC-3 \cite{KAGRA:2021duu}). With our current framework in hand, these are all straightforwards to accommodate. Such improvements will help to pin down the astrophysical parts of the analysis, lifting degeneracies with the cosmological and beyond-GR parameters. We have set the stage for improved dark siren tests of gravity, and eagerly anticipate the forthcoming GW and galaxy data.

\begin{acknowledgments}
The authors would like to thank Benoit Revenu and Christos Karathanasis for assistance with GW injection computations. A.C. is supported by a PhD grant from the Chinese Scholarship Council (grant no.202008060014), and the STSM GravNet grant during his visit to the NBI. The research of R.G. was supported by ERC Starting Grant
\textit{SHADE} 949572 and STFC grant ST/V005634/1. T.B. is supported by ERC Starting Grant \textit{SHADE} (grant no.~StG 949572) and a Royal Society University Research Fellowship (grant no.~URF$\backslash$R1$\backslash$180009). This material is based upon work supported by NSF's LIGO Laboratory, which is a major facility fully funded by the National Science Foundation. 
The authors are grateful for computational resources provided by the LIGO Laboratory and supported by  National Science Foundation Grants PHY-0757058 and PHY-0823459.
\end{acknowledgments}

\appendix

\section{Corner plots for GWTC-3 reanalysis} \label{sec:full_corner}

\begin{figure}[h]
    \centering
    \includegraphics[width=\textwidth]{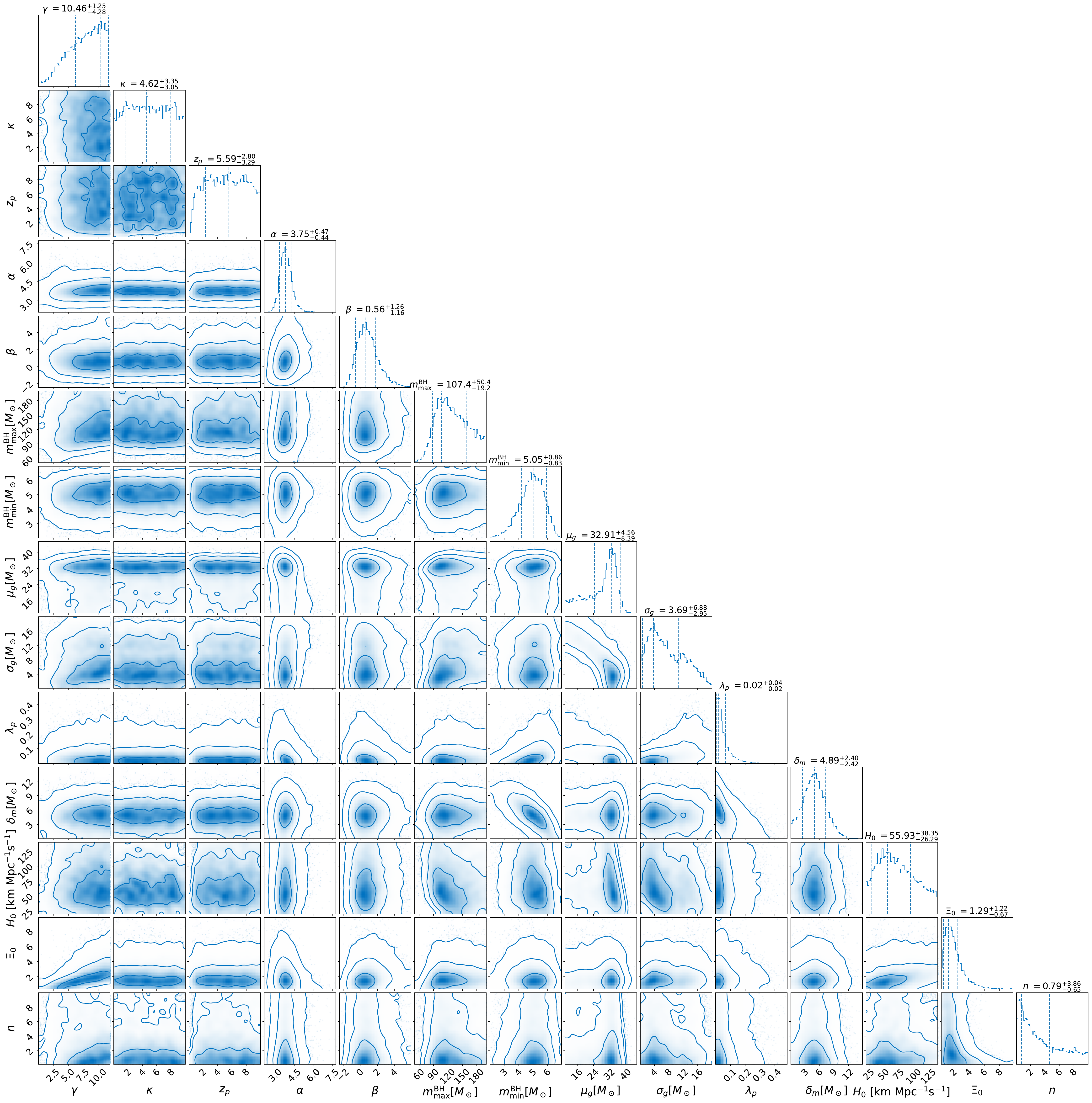}
    \caption{Full corner plot of joint posteriors for 42 BBH events from the GWTC-3 catalogue. The contours show 1, 2 and 3$\sigma$ bound for the 2D posteriors. The vertical lines in the 1D histogram plot correspond to the peak values and 1$\sigma$ bound of parameters in the title.}
    \label{fig:full_corner}
\end{figure}

\noindent We present the full corner plot for the joint posteriors obtained from the \gwcosmo sampling run with 42 BBHs in GWTC-3 in Fig. \ref{fig:full_corner}. $\gamma$, $\kappa$ and $z_p$ in the merger rate evolution model are essentially unconstrained. On the other hand, parameters for BBH population model and cosmological parameters are constrained at a better level. There exists notable degeneracy between $H_0$-$\Xi_0$, $\gamma$-$\Xi_0$, $\mu_g$-$\sigma_g$ and $m_{\rm min}^{\rm BH}$-$\sigma_m$ (see discussion in the main text).

\section{Posterior samples for GWTC-3 reanalysis}
\label{sec:posterior_samples}

\begin{figure}
    \centering
    \includegraphics[width=\textwidth]{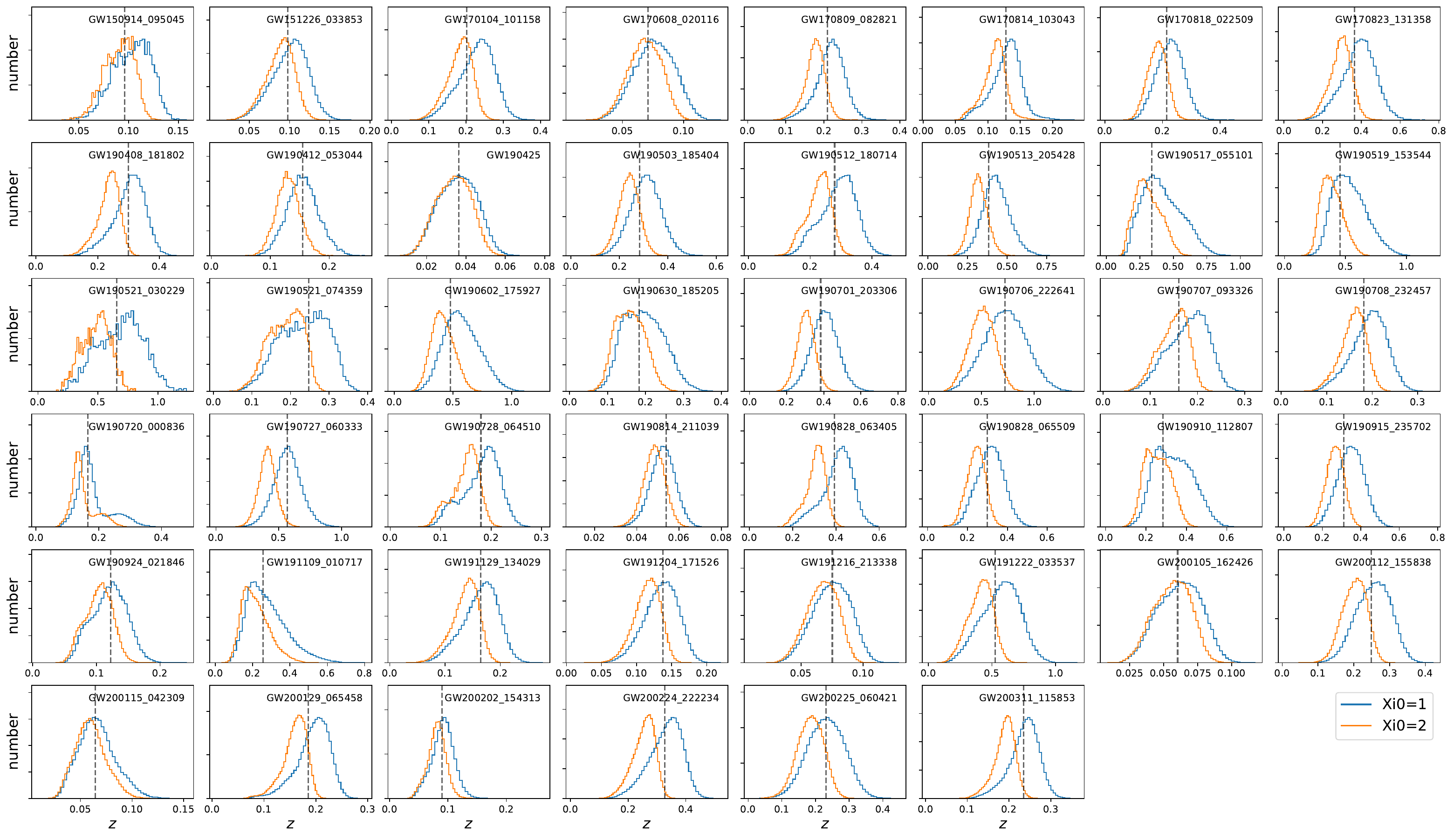}
    \caption{Posterior samples converted from GW luminosity distance to redshift for the 46 events from the GWTC-3 catalogue with different values of $\Xi_0$. The grey dashed vertical line marks the estimated redshift in GR from GWTC-3.}
    \label{fig:posterior_samples}
\end{figure}

\noindent Here we explain the three varieties of $\Xi_0$ posteriors observed from individual events in Fig~\ref{fig:Xi0_individual}. In Fig. \ref{fig:posterior_samples} we plot histograms of the redshift posterior samples for the 46 events selected from GWTC-2.1 and GWTC-3. The redshifts of posterior samples are converted from $d_L^{\rm GW}$ with two values, $\Xi_0=1$ (GR) and $\Xi_0=2$ with $H_0=70~{\rm km}~{\rm s}^{-1}{\rm Mpc}^{-1}$ and $n=1.91$. The grey dashed vertical line marks the estimated redshift in GR from GWTC-3. We can see that the posterior samples shift to lower redshift as $\Xi_0$ increases. For the events in which the peak of the samples is located 
at higher redshift than the estimated redshift, the peak crosses through to the other side of the estimated redshift when $\Xi_0$ increases. As a result, there is a peak in the likelihood of $\Xi_0$ over the prior range of $\Xi_0$, see for example GW170809\_082821 in Fig.~\ref{fig:Xi0_individual}. For the events in which the peaks of the samples for different $\Xi_0$ values are on the same side of the estimated redshift, the likelihood of $\Xi_0$ continuously decreases instead of having a peak, for example GW190910\_112807 in Fig.~\ref{fig:Xi0_individual}. For the events in which the posterior samples shift little when $\Xi_0$ increases, the likelihood is dominated by the denominator that is continuously decreasing, so the likelihood is continuously increasing, for example GW191216\_213338 in Fig.~\ref{fig:Xi0_individual}. 

\bibliographystyle{JHEP}
\bibliography{masterbib}

\end{document}